\documentclass[aps,pra,twocolumn,floatfix,groupedaddress,footinbib,notitlepage,showpacs,superscriptaddress]{revtex4-1}
\usepackage{times,bm,bbm,bbold,graphicx,graphics,amssymb,amsmath,amsfonts,dsfont,color}

\newtheorem{remark}{Remark}

\newcommand{\ket}[1]{|#1\rangle}

\newcommand{\ketbra}[2]{|#1\rangle\langle #2|}

\newcommand{\average}[1]{\left\langle#1\right\rangle}

\newcommand{\ignore}[1]{}

\usepackage[pdfstartview=FitH]{hyperref}
\hypersetup{
colorlinks=true,       	
linkcolor=blue,          	
citecolor=blue,        
filecolor=blue,      	
urlcolor=blue,           	
runcolor=blue
}

\let\oldsqrt\sqrt
\def\sqrt{\mathpalette\DHLhksqrt}
\def\DHLhksqrt#1#2{%
\setbox0=\hbox{$#1\oldsqrt{#2\,}$}\dimen0=\ht0
\advance\dimen0-0.2\ht0
\setbox2=\hbox{\vrule height\ht0 depth -\dimen0}%
{\box0\lower0.4pt\box2}}

\DeclareFontFamily{OT1}{pzc}{}
\DeclareFontShape{OT1}{pzc}{m}{it}%
              {<-> s * [1.25] pzcmi7t}{}
\DeclareMathAlphabet{\mathpzc}{OT1}{pzc}%
                                 {m}{it}

\begin{document}

\title{Correlations in quantum thermodynamics: Heat, work, and entropy production}

\author{S. Alipour}
\affiliation{School of Nano Science, Institute for Research in Fundamental Sciences (IPM), Tehran 19538, Iran}

\author{F. Benatti}
\affiliation{Department of Physics, University of Trieste, I-34151 Trieste, Italy}
\affiliation{National Institute for Nuclear Physics (INFN), Trieste Section, I-34151 Trieste, Italy}

\author{F. Bakhshinezhad}
\affiliation{Department of Physics, Sharif University of Technology, Tehran 14588, Iran}

\author{M. Afsary}
\affiliation{Department of Physics, Sharif University of Technology, Tehran 14588, Iran}

\author{S. Marcantoni}
\affiliation{Department of Physics, University of Trieste, I-34151 Trieste, Italy}
\affiliation{National Institute for Nuclear Physics (INFN), Trieste Section, I-34151 Trieste, Italy}

\author{A. T. Rezakhani}
\email{rezakhani@sharif.edu}
\affiliation{Department of Physics, Sharif University of Technology, Tehran 14588, Iran}

\begin{abstract}
We provide a characterization of energy in the form of exchanged heat and work between two interacting constituents of a closed, bipartite, correlated quantum system. By defining a binding energy we derive a consistent quantum formulation of the first law of thermodynamics, in which the role of correlations becomes evident, and this formulation reduces to the standard classical picture in relevant systems. We next discuss the emergence of the second law of thermodynamics under certain---but fairly general---conditions such as the Markovian assumption. We illustrate the role of correlations and interactions in thermodynamics through two examples.
\end{abstract}

\pacs{05.30-d , 05.70.-a , 03.65.Ud}
\maketitle

\section{Introduction}

Investigating the consistency of thermodynamics, as a successful classical theory of macroscopic physical systems, with quantum mechanics, as a fundamental theory of the underlying microscopic systems, is still an open problem and the subject of extensive recent research \cite{Book:Gemmer}. Defining quantum mechanical counterparts of the classical concepts of ``heat'' and ``work'' and describing the mechanisms underlying their exchange among microscopic quantum systems are the first steps towards this goal. From this point of view, studying the thermodynamic role of quantum mechanical features such as nonclassical correlations is of paramount importance.

There exist various approaches to defining heat and work \textit{microscopically}. One widely-used definition has been proposed in Ref.~\cite{Alicki-79}, where  work exchange is due to an external periodic driving incorporated in a time-dependent Hamiltonian, while heat is absorbed or released because of interaction with an ambient environment. In order to show that also in time-independent Hamiltonian systems work can be extracted, other approaches have been proposed, such as using another quantum system as a ``work storage'' \cite{Popescu}. These proposed mechanisms, however, are not entirely consistent with each other. Indeed, the role of correlations is usually not considered in these approaches since the focus is mainly on one of the two parties of compound quantum systems, whereas a comprehensive study of thermodynamic behaviors should consider all effects concerning systems involved in nonequilibrium thermodynamic processes.

Here we revisit the first and the second laws of thermodynamics in a closed (conservative) quantum system comprising two interacting parties, one denoted by $S$ as the thermodynamic system of interest and the other denoted by $B$ playing the role of its environment, and reformulate these laws in a way that clearly exposes the role played by $SB$ correlations. In order to do so we introduce the notion of  ``binding energy" of two interacting quantum systems which, together with the internal energies of the two parties, provides the internal energy of the whole system.

The definitions of heat and work that we use are similar to those in Ref.~\cite{Alicki-79}; however, unlike there, we show that, in general, nonequilibrium thermodynamic processes affecting a system $S$ involve work exchange with $B$ without the need for an external driving represented by a time-dependent parameter in the system Hamiltonian, but merely because of the interactions between $S$ and $B$. Besides, we explicitly show that correlations do not play any role in work exchange, while they do play an unavoidable role in heat transfer between $S$ and $B$. Furthermore, this latter process does not necessarily need $B$ to be treated as an environment weakly coupled to $S$, thus it need not be expected to evolve in time according to a dissipative Markovian (Lindbladian) dynamics.

As a preliminary and necessary step toward investigating heat and work exchanges between two interacting systems $S$ and $B$, one needs to unambiguously assign to the two parties a percentage of the interaction energy depending on the state of the compound system. However, due to $SB$ correlations, there will always be part of the interaction energy that belongs to both $S$ and $B$ together. In thermodynamic terms, extracting this part of the energy would require accessibility of the total system. Thus, we distinguish three contributions to the total internal energy of $SB$: one  accessible only through $S$, the other one only through $B$, and the last one only through $SB$ (as a whole) via the $SB$ correlations. We call this latter contribution to the internal energy the \textit{binding energy}. Certainly, although (in the case of time-independent total Hamiltonian) the total internal energy remains constant in time, that of either $S$ or $B$ varies because they interact and thus exchange work and heat. 

In a recent publication \cite{Castro}, the internal energy of an open quantum system has been defined as the energy which is accessible through \textit{measurements} in a fixed ``local effective measurement basis'' \cite{Weimer}, and the definitions of work and heat suggested by considering the ability of the energy changes in altering the von Neumann entropy; heat is the energy flux that may change the entropy but work is the part of the energy change that keeps entropy intact. In contrast, in our formalism the internal energy associated with each subsystem is defined as the energy which is locally accessible in each individual subsystem by means of arbitrary local measurements. Although it has been known in the literature that correlations play a role in heat exchange, this fact has not been shown explicitly thus far. In the following, we provide explicit relations that exhibit the role played by correlations in heat, work, and entropy exchange between constituents of a bipartite system.

The structure of this paper is as follows. In Sec.~\ref{sec:1st-law} we lay out the foundation to define basic thermodynamic properties such as heat and work, and show in Sec. \ref{sec:exchange} a first law governing their mutual transformations. Section \ref{sec:2nd-law} deals with finite and infinitesimal versions of a possible formulation of the second law of thermodynamics. We illustrate our formalism through two examples in Sec.~\ref{sec:examples}. The paper is concluded by a summary in Sec. \ref{sec:summary}.  

\section{First law of thermodynamics in the presence of interactions and correlations}
\label{sec:1st-law}

We consider a closed quantum system $SB$ consisting of two interacting quantum systems: the system of interest $S$ and its \textit{bath} or \textit{environment} $B$---with no restrictions on the dimensionality of $S$ and $B$. The state of $SB$ is described by the density matrix $\varrho_{SB}(\tau)$, evolving under a total time-independent Hamiltonian
\begin{equation}
H_{\mathrm{tot}}=H_S+H_B+H_{\mathrm{int}}.
\label{tot-Ham}
\end{equation}
The \textit{internal energy} of the total system is the mean value of the total Hamiltonian with respect to the time-evolving state, namely $\mathds{U}_{\mathrm{tot}}=\mathrm{Tr}[\varrho_{SB}(\tau) H_{\mathrm{tot}}]$, and is thus constant in time since the dynamics of the total system is governed by the Schr\"odinger equation
\begin{equation}
\label{timevSB}
\mathrm{d}\varrho_{SB}(\tau)=-i[H_{\mathrm{tot}},\varrho_{SB}(\tau)]\,\mathrm{d}\tau.
\end{equation}
We assume $\hbar\equiv 1$ throughout the paper.

In thermodynamics, infinitesimal variations of the internal energy of a system occur because of infinitesimal exchanges of \textit{heat} $\mathds{Q}$ and/or \textit{work} $\mathds{W}$ between the system and the environment. The quantum mechanical counterparts of infinitesimal heat and work exchanges in a system with state $\varrho(\tau)$ and time-dependent Hamiltonian $H(\tau)$ are given by \cite{Balian}
\begin{align}
\mathrm{d}\mathds{Q}(\tau)&=\mathrm{Tr}\left[\mathrm{d}\varrho(\tau)H(\tau)\right],
\label{heat}\\
\mathrm{d}\mathds{W}(\tau)&=\mathrm{Tr}\left[\varrho(\tau)\mathrm{d}{H}(\tau)\right],
\label{work}
\end{align}
where ``$\mathrm{d}$" denotes a time differential, while $\mathrm{d}\mathds Q$ and $\mathrm{d}\mathds W$ are in general inexact differentials. With these definitions, we have the following quantum version of the first law of thermodynamics for the internal energy $\mathds{U}(\tau)=\mathrm{Tr}\left[H(\tau)\varrho(\tau)\right]$:
\begin{equation}
\label{first-law}
\mathrm{d}\mathds{U}(\tau)=\mathrm{d}\mathds{Q}(\tau)+\mathrm{d}\mathds{W}(\tau).
\end{equation}

In the case of a compound, isolated system $SB$, the states of the constituent subsystems are obtained by partial tracing $\varrho_{S,B}(\tau)=\mathrm{Tr}_{B,S}\left[\varrho_{SB}(\tau)\right]$; thus from Eq.~(\ref{timevSB}) we have 
\begin{align}
\mathrm{d}\varrho_S(\tau)=&-i[H_S,\varrho_S(\tau)]\mathrm{d}\tau -i\mathrm{Tr}_B\left[H_{\mathrm{int}},\varrho_{SB}(\tau)\right]\mathrm{d}\tau,
\label{PT1}
\\
\mathrm{d}\varrho_B(\tau)=&-i[H_B,\varrho_B(\tau)]\mathrm{d}\tau -i\mathrm{Tr}_S\left[H_{\mathrm{int}},\varrho_{SB}(\tau)\right]\mathrm{d}\tau.
\label{PT2}
\end{align}
When there are \textit{correlations} between the system and the environment, we can write
\begin{equation}
\varrho_{SB}(\tau)=\varrho_S(\tau)\otimes\varrho_B(\tau)+\chi(\tau),
\label{corr}
\end{equation}
where $\chi$ measures all correlations (classical or quantum). Replacing this decomposition into Eqs.~(\ref{PT1}) and (\ref{PT2}) yields
\begin{align}
\mathrm{d}\varrho_S(\tau)=&-i[H'_S(\tau),\varrho_S(\tau)]\mathrm{d}\tau -i\mathrm{Tr}_B\left[H_{\mathrm{int}},\chi(\tau)\right]\,\mathrm{d}\tau ,\label{PT3}
\\
\mathrm{d}\varrho_B(\tau)=&-i[H'_B(\tau),\varrho_B(\tau)]\mathrm{d}\tau -i\mathrm{Tr}_S\left[H_{\mathrm{int}},\chi(\tau)\right]\,\mathrm{d}\tau,\label{PT4}
\end{align}
where we have modified the Hamiltonians as
\begin{equation}
H'_{S,B}(\tau)=H_{S,B}+\mathrm{Tr}_{B,S}\left[\varrho_{B,S}(\tau)H_{\mathrm{int}}\right].
\end{equation}
The last term might be reminiscent of \textit{Lamb shift} corrections in Markovian dynamics \cite{Book:Breuer}. We can rewrite Eq.~(\ref{tot-Ham}) as 
\begin{equation}
H_{\mathrm{tot}}=H'_{S}(\tau)+H'_{B}(\tau)+H'_{\mathrm{int}}(\tau),
\label{binden0c}
\end{equation}
where
\begin{equation}
H'_{\mathrm{int}}(\tau)=H_{\mathrm{int}}-\mathrm{Tr}_{S}\left[\varrho_{S}(\tau)H_{\mathrm{int}}\right]
-\mathrm{Tr}_{B}\left[\varrho_{B}(\tau)H_{\mathrm{int}}\right].
\end{equation}

Since the interaction Hamiltonian is nonlocal, it seems physically reasonable to expect that it is not accessible by local means. This is mathematically enforced by requesting that the mean values of $H'_{\mathrm{int}}(\tau)$ with respect to the local states vanish. However, we have
\begin{align}
\mathrm{Tr}_{S}[\varrho_S(\tau)H'_{\mathrm{int}}(\tau)]&=\mathrm{Tr}_{B}[\varrho_B(\tau)H'_{\mathrm{int}}(\tau)]\nonumber\\
&=-\mathrm{Tr}[\varrho_S(\tau)\otimes\varrho_B(\tau)H_{\mathrm{int}}]. \label{scalar}
\end{align}
To remedy this, we simply need to compensate for the nonzero scalar contribution of Eq.~(\ref{scalar}) by distributing it over the system and environment Hamiltonians through the real (but not necessarily positive) auxiliary parameters $\alpha_{S}$ and $\alpha_{B}=1-\alpha_{S}$ and hence defining \textit{effective} Hamiltonians
\begin{align}
H_{S}^{(\mathrm{eff})}(\tau)&=H'_{S}(\tau)-\alpha_{S}\mathrm{Tr}\left[\varrho_S(\tau)\otimes \varrho_B(\tau)H_{\mathrm{int}}\right],
\label{binden2}\\
H_{B}^{(\mathrm{eff})}(\tau)&=H'_{B}(\tau)-\alpha_{B}\mathrm{Tr}\left[\varrho_S(\tau)\otimes \varrho_B(\tau)H_{\mathrm{int}}\right].
\label{binden22}
\end{align}
Accordingly, $H'_{\mathrm{int}}$ in Eq.~\eqref{binden0c} can be replaced with an effective interaction Hamiltonian
\begin{align}
H_{\mathrm{int}}^{(\mathrm{eff})}(\tau)=H_{\mathrm{tot}}-H_{S}^{(\mathrm{eff})}(\tau)-H_{B}^{(\mathrm{eff})}(\tau). 
\label{binden3}
\end{align}
Note that Eqs.~(\ref{PT3}) and (\ref{PT4}) remain valid where $H'_{S,B}(\tau)$ are replaced with $H^{(\mathrm{eff})}_{S,B}(\tau)$, and we have
\begin{equation}
\mathrm{Tr}_{S}[\varrho_S(\tau)\,H_{\mathrm{int}}^{(\mathrm{eff})}(\tau)]=\mathrm{Tr}_{B}[\varrho_B(\tau)H_{\mathrm{int}}^{(\mathrm{eff})}(\tau)]=0.
\end{equation}

By defining the \textit{internal} energies of the constituent systems through the effective Hamiltonians
\begin{equation}
\mathds{U}_{S,B}(\tau)=\mathrm{Tr}[\varrho_{S,B}(\tau)H_{S,B}^{(\mathrm{eff})}(\tau)],
\label{Us}
\end{equation}
an energy contribution remains, called \textit{binding} energy, which can be naturally attributed to correlations $\chi$ as
\begin{equation}
\label{binden1}
\mathds{U}_{\chi}(\tau)=\mathrm{Tr}[\chi(\tau)H_{\mathrm{int}}^{(\mathrm{eff})}(\tau)],
\end{equation}
such that
\begin{equation}
\label{tot-ener}
\mathds{U}_{\mathrm{tot}}=\mathds{U}_{S}(\tau)+\mathds{U}_{B}(\tau) + \mathds{U}_{\chi}(\tau).
\end{equation}

Note that only if the interaction and correlations between the two systems were negligible (which is usually assumed in classical thermodynamics)---namely only if $H_{\mathrm{int}}\approx 0$ and $\varrho_{SB}(\tau)\approx \varrho_S(\tau)\otimes\varrho_B(\tau)$ at all times $\tau$---we would have $\mathds{U}_{\mathrm{tot}}\approx\mathds{U}_S(\tau)+\mathds{U}_B(\tau)$ and the internal energy be additive.

\section{Exchange of heat and work between two interacting systems}
\label{sec:exchange}

When two thermodynamical systems are independent, namely uncorrelated and noninteracting, the amount of the transferred heat/work from one system is equivalent to the heat/work received by the other system. In our context, the existence of interactions and correlations between $S$ and $B$ alters this picture for heat exchange, but not for the exchanged work. Indeed, inserting Eq.~\eqref{binden2} into Eq.~\eqref{work}, the infinitesimal works performed by $S$ and $B$ are obtained as
\begin{align}
\label{workSB1}
\mathrm{d}\mathds{W}_S(\tau)=&\alpha_B\mathrm{Tr}\left[\varrho_S(\tau)\otimes\mathrm{d}\varrho_B(\tau)H_{\mathrm{int}}\right]\\
&
-\alpha_S\mathrm{Tr}\left[\mathrm{d}\varrho_S(\tau)\otimes\varrho_B(\tau)H_{\mathrm{int}}\right]=-\mathrm{d}\mathds{W}_B(\tau).\nonumber
\end{align}
The work absorbed or released by system $S$, respectively $B$, each depends on the scalars $\alpha_{S,B}$, which is reminiscent to the non-gauge-invariance feature of work \cite{Hanggi-work}. Nevertheless, we always have
\begin{equation}
\label{workSB2}
\mathrm{d}\mathds{W}_S(\tau)+\mathrm{d}\mathds{W}_B(\tau)=0.
\end{equation}
Note that, unlike in Refs.~\cite{Alicki-79,Alicki} where it is the time dependence of the Hamiltonian through an external driving parameter that leads to work exchange, here the work exchange follows from the time dependence of the effective Hamiltonians which include time-dependent Lamb-shift-like corrections in Eq.~\eqref{binden3}. As a consequence, our formalism features that work exchange between two interacting subsystems is allowed even without an external driving.

Unlike the infinitesimal work $\mathrm{d}\mathds{W}_S(\tau)$, the scalar parameters $\alpha_{S,B}$ do not contribute to the infinitesimal heat exchanges because $\mathrm{Tr}\left[\mathrm{d}\varrho(\tau)\right]=0$. In particular, inserting Eqs.~\eqref{binden2} and \eqref{binden22} into Eq.~\eqref{heat} yields
\begin{align}
\label{heatSB1}
\mathrm{d}\mathds{Q}_S(\tau)&=-i\mathrm{Tr}\left[\chi(\tau)\big[H^{(\mathrm{eff})}_S(\tau),H_{\mathrm{int}}\big]\right]\mathrm{d}\tau,\\
\mathrm{d}\mathds{Q}_B(\tau)&=-i\mathrm{Tr}\left[\chi(\tau)\big[H^{(\mathrm{eff})}_B(\tau),H_{\mathrm{int}}\big]\right]\mathrm{d}\tau.
\end{align}
In addition, differentiating Eq.~\eqref{tot-ener} and using Eq.~\eqref{first-law} together with Eq.~\eqref{workSB2}, one obtains that
\begin{equation}
\label{heatSB2}
\mathrm{d}\mathds{Q}_S(\tau)+\mathrm{d}\mathds{Q}_B(\tau)=-\mathrm{d}\mathds{U}_\chi(\tau).
\end{equation}
Hence, in our setting the binding energy is completely of the heat type. This can also be seen from the relations $\mathrm{d}\mathds{U}_{\chi}(\tau)=\mathrm{Tr}[\mathrm{d}\chi(\tau)H_{\mathrm{int}}^{(\mathrm{eff})}(\tau)]$ and $\mathrm{Tr}[\chi(\tau)\mathrm{d}H_{\mathrm{int}}^{(\mathrm{eff})}(\tau)]=0$, which show that changes in the binding energy can only come from changes in the state correlations. Indeed, the heat balance equation (\ref{heatSB2}) shows that (i) heat transfer is only due to interactions and correlations within the total system, in agreement with the result of Refs.~\cite{Schroder-Mahler,Weimer}, and (ii) that heat passing from one system to the other is paid for by varying the $SB$ correlations that thus behave like a \textit{heat storage}. That is, if correlations do not change and $\mathrm{d}\chi(\tau)=0$, then $\mathrm{d}\mathds{U}_{\chi}=0$; whence $~\mathrm{d} \mathds{Q}_S=-\mathrm{d}\mathds{Q}_B$, in agreement with the standard textbook definition of ``heat" in classical systems wherein no or only a negligibly-weak interaction between the system and the environment is assumed \cite{Book:thermodynamics}.

\begin{remark} \label{r1}
In our derivations thus far we have assumed that the Hamiltonian of the total system $SB$ is time-independent. If we relax this condition and allow a time dependence (e.g., due to the action of some external agent on the total system), part of our relations will be modified as follows:
\begin{align}
&\mathrm{d}\mathds{U}_{\mathrm{tot}}(\tau)= \mathrm{d}\mathds{U}_{S}(\tau) + \mathrm{d}\mathds{U}_{B}(\tau)+ \mathrm{d}\mathds{U}_{\chi}(\tau),\\
&\mathrm{d}\mathds{Q}_{S}(\tau) + \mathrm{d}\mathds{Q}_{B}(\tau) = -\mathrm{d}\mathds{Q}_{\chi}(\tau),\\
&\mathrm{d}\mathds{W}_{S}(\tau) + \mathrm{d}\mathds{W}_{B}(\tau) = \mathrm{Tr}[\varrho_{S}(\tau) \otimes \varrho_{B}(\tau)\,\mathrm{d}H_{\mathrm{int}}(\tau)], \label{last-t}
\end{align}
 where $\mathrm{d}\mathds{Q}_{\chi}(\tau)= \mathrm{Tr}[\mathrm{d}\chi(\tau)\,H_{\mathrm{int}}(\tau)]$. Equation (\ref{last-t}) indicates that even in the time-dependent case correlations do not contribute in the exchange of work between the system and the environment. Since the dynamics of the compound system $SB$ is generated by $H_{\mathrm{tot}}(\tau)$, it follows that $\mathrm{Tr}[\mathrm{d}\varrho_{SB}(\tau)H_{\mathrm{tot}}(\tau)]=0$ (i.e., the total system is thermally isolated), so that there are no heat exchanges and the only possibility for $SB$ is to perform work because of the external driving due to the rest of the universe.
 
\end{remark}

\section{Second law of thermodynamics}
\label{sec:2nd-law}

According to the second law of thermodynamics, the entropy of a macroscopic closed system which is \textit{thermally isolated} (in thermodynamics terminology) can only remain constant or increase in time \cite{Landau,Book:thermodynamics}. However, the second law is not necessarily valid in nonequilibrium microscopic or even macroscopic systems \cite{Evans,Crooks,Jarzynski,An:Nature-Phys,sl-1,sl-2}. 

In the following we demonstrate the possible emergence of the second law of thermodynamics and the important role of system-bath correlations in this microscopic context.

In the case of a compound system $SB$, the subadditivity of the von Neumann entropy \cite{book:Nielsen} (we set $\kappa_{\mathrm{B}}\equiv 1$ for the Boltzmann constant throughout the paper)
\begin{equation}
\label{vNe}
\mathds{S}(\tau)=-\mathrm{Tr}\left[\varrho(\tau)\log\varrho(\tau)\right],
\end{equation}
implies that the \textit{mutual information}
\begin{equation}
\mathds{S}_{\chi}(\tau) =\mathds{S}_S (\tau)+\mathds{S}_B(\tau)-\mathds{S}_{SB}(\tau)
\label{S}
\end{equation}
is always nonnegative. Mutual information characterizes the amount of total correlations (both classical and quantum) shared by the two subsystems $S$ and $B$ \cite{wolf-PRL,Hall}. Intuitively, if the correlations between $S$ and $B$ increases, $\mathds{S}_{\chi}$ becomes larger.

Since we have assumed that the total system $SB$ is closed, it evolves unitarily and its von Neumann entropy $\mathds{S}_{SB}(\tau)$ does not change in time (even if its Hamiltonian depends on time). Hence, differentiating Eq.~\eqref{S} yields
\begin{equation}
\mathrm{d}\mathds{S}_S (\tau)+\mathrm{d}\mathds{S}_B(\tau)=\mathrm{d}\mathds{S}_{\chi}(\tau).
\label{ds}
\end{equation}
Integrating both sides of this equation in the time interval $[0,\tau]$, with the assumption that the initial state of $SB$ is uncorrelated (i.e., $\mathds{S}_\chi(0)=0$), leads to
\begin{equation}
\Delta\mathds{S}_S+\Delta\mathds{S}_B=\mathds{S}_{\chi}(\tau)\geqslant 0,
\label{ent}
\end{equation}
as obtained in Ref.~\cite{Reeb-Wolf}. This relation states that, as long as one observes subsystems $S$ and $B$ locally and their initial state is without any correlations, the sum of the total variations of the entropies of $S$ and $B$ is always nonnegative. One can consider this property as a form of the second law of thermodynamics for the compound system $SB$. 

Unlike in equilibrium thermodynamics, in a general nonequilibrium system ``temperature" is not a well-defined quantity (see, e.g., Refs.~\cite{nT,FDT} for some recent discussions). However, at fixed ``volume" ($V$) and ``number of particles" ($N$), one can introduce a time-dependent \textit{pseudo-temperature} by means of the internal energy and the von Neumann entropy  through
\begin{equation}
\frac{1}{T(\tau)}:=\frac{\mathrm{d}\mathds{S}(\tau)}{\mathrm{d}\mathds{U}(\tau)},
\label{T-def}
\end{equation}
which is somewhat reminiscent of the standard, equilibrium definition $1/T=(\partial \mathds{S}/\partial \mathds{U})_{N,V}$. 

\begin{remark} \label{rem:2}
In generic quantum systems, it is not always clear how to define $V$ and $N$ (or other relevant thermodynamic properties) in general quantum systems. Additionally, in thermodynamic equilibrium we deal with the partial derivative $(\partial \mathds{S}/\partial \mathds{U})_{N,V}$ rather than the ratio of two total derivatives ($(\mathrm{d}\mathds{S}/\mathrm{d}\tau)/(\mathrm{d}\mathds{U}/\mathrm{d}\tau)$), which can be different quantities. Noting Eqs.~(\ref{binden2}) and (\ref{binden22}), the free parameter $\alpha_{S}$ (and $\alpha_{B}$) would also appear in the pseudo-temperature. In general then, one should not expect that the pseudo-temperature necessarily have definite relation with the equilibrium temperature, unless under certain conditions. Later in the examples we show explicitly how in special cases the pseudo-temperature may relate to the equilibrium temperature by appropriately fixing the scalar $\alpha_{S}$ through thermodynamic properties of the system in question. 
\end{remark}

Adopting the concept of pseudo-temperature, one can associate (time-dependent) pseudo-temperatures $T_{S,B}(\tau)$ with subsystems $S$ and $B$ and also a pseudo-temperature $T_\chi(\tau)$ with the binding energy. As a result, inserting Eq.~\eqref{T-def} into Eq.~\eqref{ds} gives
\begin{equation}
\frac{\mathrm{d}\mathds{U}_S(\tau)}{T_S(\tau)}+\frac{\mathrm{d}\mathds{U}_B(\tau)}{T_B(\tau)}=\frac{\mathrm{d}\mathds{U}_{\chi}(\tau)}{T_{\chi}(\tau)}.
\label{T}
\end{equation}
It then follows that, when $\mathrm{d}\mathds{U}_{\chi}(\tau)=0$ but $T_\chi(\tau)\neq 0$, since $\mathrm{d}\mathds{U}_{S}(\tau)=-\mathrm{d}\mathds{U}_{B}(\tau)$ [Eq.~(\ref{tot-ener})], the two subsystems must have the same instantaneous pseudo-temperature: $T_S(\tau)=T_B(\tau)$. Another possibility is when $\mathrm{d}\mathds{U}_{\chi}(\tau)=0$ and $T_\chi(\tau)= 0$ such that $\mathrm{d}\mathds{S}_\chi(\tau)\neq 0$, whence 
\begin{equation}
\mathrm{d}\mathds{U}_B(\tau)=\frac{T_S(\tau)T_B(\tau)}{T_S(\tau)-T_B(\tau)}\, \mathrm{d}\mathds{S}_\chi(\tau).
\label{important}
\end{equation} 
We remark that a somewhat similar result in Ref.~\cite{Lloyd-et al} is akin to our general expression in Eq. \eqref{T} (of course here with pseudo-temperature instead of equilibrium temperature). This interesting relation yet agin indicated the role of correlations; for energy transport to the bath (and similarly to the system) correlations are necessary, where in turn development of correlations ensues from interaction.

\begin{remark}
Under the same conditions, when the total Hamiltonian is time-dependent, Eq.~(\ref{important}) is modified to
\begin{equation}
\mathrm{d}\mathds{U}_B(\tau)=\frac{T_S(\tau)T_B(\tau)}{T_S(\tau)-T_B(\tau)}\left(\mathrm{d}\mathds{S}_{\chi}(\tau)-\frac{\mathrm{d}\mathds{U}_{\mathrm{tot}}(\tau)}{T_{S}(\tau)}\right).
\label{important'}
\end{equation}
\end{remark}

Furthermore, from Eqs.~\eqref{first-law} and~\eqref{T-def}, it follows that
\begin{equation}
\mathrm{d}\mathds{S}_{S,B}(\tau)=\frac{\mathrm{d}\mathds{Q}_{S,B}(\tau)}{T_{S,B}(\tau)}+\frac{\mathrm{d}\mathds{W}_{S,B}(\tau)}{T_{S,B}(\tau)}.
\label{entprod1}
\end{equation}
Formally, the difference between the total variations of the entropies and the contributions coming from the heat exchanges are given by the contributions due to the work exchanges, 
\begin{equation}
\label{entbal}
\mathrm{d}\widetilde{\Sigma}_{S,B}(\tau):=\mathrm{d}S_{S,B}(\tau)-\frac{\mathrm{d} \mathds{Q}_{S,B}(\tau)}{T_{S,B}(\tau)}=\frac{\mathrm{d} \mathds{W}_{S,B}(\tau)}{T_{S,B}(\tau)}.
\end{equation}
The quantity $\mathrm{d}\widetilde{\Sigma}$ resembles the infinitesimal \textit{internal entropy production} as defined in Ref.~\cite{Alicki-79}, where the case of an externally driven system $S$ has been discussed which is weakly coupled to a conservative heat bath $B$ inducing a dissipative dynamics \cite{Lindblad,Gorini,Book:Breuer,Book:Carmichael} (in general explicitly time-dependent) of the Lindblad type. In this particular context, the infinitesimal entropy production is modified as the difference between the variation of the entropy $\mathds{S}_S(\tau)$ and the entropy flux into or out of the system associated to the heat flux $\mathrm{d}\mathds{Q}_S(\tau)$ divided by the (initial) temperature of the bath $T$ (rather than $T_{S}(\tau)$ as in Eq.~(\ref{entprod1})),
\begin{equation}
\label{iepA}
{\mathrm d}\Sigma_S(\tau):=\mathrm{d}\mathds{S}_S(\tau)-\frac{\mathrm{d}\mathds{Q}_S(\tau)}{T}.
\end{equation}

This expression can be interpreted as an internal entropy production for system $S$ and its nonnegativity, ${\mathrm d}\Sigma(\tau)\geqslant0$, can be considered as an infinitesimal expression of the second law of thermodynamics. If the time evolution $\varrho_S(0)\mapsto\varrho_S (\tau)$ is given by a Lindblad-type generator $\mathpzc{L}_\tau$ that preserves the instantaneous Gibbs state $\varrho_S^{\beta}(\tau)= \exp(-\beta H_S(\tau))/Z_\beta(\tau)$, with $\beta=1/T$ and $Z_{\beta}(\tau)=\mathrm{Tr}\left[{\rm e}^{-\beta H_S(\tau)}\right]$, namely $\mathpzc{L}_{\tau}[\varrho_S^\beta(\tau)]= 0$, then one can recast the infinitesimal entropy production (\ref{iepA}) as
\begin{equation}
\mathrm{d}\Sigma_S(\tau)=-\mathrm{Tr}\left[\mathpzc{L}_{\tau}[\varrho_{S}(\tau)]
\left( \log\varrho_{S}(\tau) - \log\varrho^{\beta}_{S}(\tau)\right)\right]\mathrm{d}\tau.
\end{equation}
If the generator $\mathpzc{L}_{\tau}$ is of the Lindblad form, for each fixed $\tau\geqslant 0$, the maps $\mathpzc{E}_{s}=\mathrm{e}^{s \mathpzc{L}_{\tau}}$ form---with respect to the nonnegative parameter $s$---a semigroup of completely-positive and trace preserving family of maps \cite{book:Nielsen}. Since $\mathpzc{E}_{s}[\varrho_{S}^{\beta}(\tau)]=\varrho_S^{\beta}(\tau)$ and the relative entropy $D\big(\mathpzc{E}_{s}[\varrho_{S}(\tau)] \vert\hskip-.2mm\vert \varrho^\beta_{S}(\tau)\big)$ is a monotonically decreasing function of $s$ \cite{wehrl78}, one obtains the following infinitesimal quantum version of the second law of thermodynamics:
\begin{equation}
\left.\mathrm{d}\Sigma_S(\tau)=-\frac{\mathrm{d}}{\mathrm{d}s}D\big(\mathpzc{E}_{s}[\varrho_{S}(\tau)] \vert\hskip-.2mm\vert \varrho^\beta_{S}(\tau)\big)\right|_{s =0}\mathrm{d}\tau \geqslant0.
\label{posentprod}
\end{equation}

\begin{remark}
A simpler physical context is provided when there is no external driving for system $S$, namely its Hamiltonian $H_{S}(\tau)=H_S$ is time-independent, and so are the Lindblad generator $\mathpzc{L}_\tau=\mathpzc{L}$ of its dissipative dynamics $\varrho_S(0)\mapsto\varrho_S (\tau)$ and the Gibbs state $\varrho_S^\beta$ (such that $\mathpzc{L}[\varrho_S^\beta]=0$). In such a case, the proof of the positivity of the entropy production follows from Eq.~\eqref{iepA} becoming
\begin{equation}
\mathrm{d}\Sigma_S(\tau)=-\mathrm{d}D\big(\varrho_S(\tau)\vert\hskip-.2mm\vert\varrho_S^\beta\big)
\end{equation}
and from the monotonicity of the relative entropy under completely-positive, trace preserving dynamics.
\end{remark}

In the finite expression of the second law of thermodynamics (which follows from Eq.~\eqref{ent} in the absence of initial correlations between $S$ and $B$), the heat bath $B$ is taken explicitly and directly into account (though the term $\Delta \mathds{S}_{B}$). Rather, in the infinitesimal expression \eqref{posentprod}, the heat bath is indirectly accounted for by the fact that (i) the heat exchange occurs at the bath temperature, and (ii) that the dissipative reduced dynamics of system $S$ is determined by the bath in the weak-coupling limit. 

Notwithstanding these fundamental physical differences, it is still interesting to study to which extent the thermodynamical inequality $\mathrm{d}\Sigma_S(\tau)\geqslant0$ can be related to the behavior of $\mathrm{d}\widetilde{\Sigma}_{S,B}(\tau)$ in Eq.~\eqref{entbal}. It is evident that $\mathrm{d}\widetilde{\Sigma}_{S,B}(\tau)$ cannot be both strictly positive in general. For example, in the case of the same instantaneous pseudo-temperatures, as when $\mathrm{d}\mathds{U}_\chi(\tau)=0$ and $T_\chi(\tau)\neq 0$, from
\begin{equation}
\label{infbalance}
\mathrm{d}\widetilde{\Sigma}_S(\tau)+\mathrm{d}\widetilde{\Sigma}_B(\tau)=
\frac{\mathrm{d}\mathds{W}_S(\tau)}{T_S(\tau)}+\frac{\mathrm{d} \mathds{W}_B(\tau)}{T_B(\tau)}
\end{equation}
we obtain $\mathrm{d}\widetilde{\Sigma}_S(\tau)=-\mathrm{d}\widetilde{\Sigma}_B(\tau)$ for $\mathrm{d} \mathds{W}_S(\tau)=-\mathrm{d} \mathds{W}_B(\tau)$. In general, it is not true that the finite variation
\begin{equation}
\label{finbalance}
\Delta\widetilde{\Sigma}_S(\tau)+\Delta\widetilde{\Sigma}_B(\tau)=
\int_0^\tau\mathrm{d}s\,\frac{\mathrm{d}\mathds{W}_S(s)}{\mathrm{d}s}\left(\frac{1}{T_S(s)}-\frac{1}{T_B(s)}\right)
\end{equation}
becomes nonnegative in the absence of initial correlations between $S$ and $B$---unlike the case for the finite variations of the von Neumann entropies of the reduced states $\varrho_{S,B}(\tau)$.

One can argue that the infinitesimal quantities $\mathrm{d}\widetilde{\Sigma}_{S,B}(\tau)$ do not generically behave as expected from true thermodynamic quantities because the instantaneous pseudo-temperatures do not behave themselves as thermodynamic temperatures. This, however, does not exclude that, under certain conditions, proper thermodynamic patterns might emerge.

To alleviate the above situation, we can discern a better motivated notion of temperature by appealing to analogy with standard thermodynamics. In \textit{classical} thermodynamics the relation
\begin{equation}
\mathrm{d}\mathds{S}= \frac{1}{T}\mathrm{d}\mathds{Q}
\end{equation}
holds for a system undergoing a quasistatic reversible transformation, whereas for a nonequilibrium process there is an extra term corresponding to the internal entropy production $\Sigma$,
\begin{equation}
\mathrm{d}\mathds{S}= \frac{1}{T}\mathrm{d}\mathds{Q} + \mathrm{d}\Sigma.
\label{entro_prod}
\end{equation}
In this case the ``temperature" is fixed by the external environment (bath) which is supposed to exchange heat always quasistatically (because of its short relaxation times), \textit{without changing its temperature}. In our formalism, however, we treat the system and bath similarly. Thus we can extend Eq. \eqref{entro_prod} and identify an \textit{extended} temperature and an entropy production for both system and bath and see how they compare at long times with expected thermodynamic temperatures. One way to do so is to explicitly compute $\mathrm{d}\mathds{S}$ and $\mathrm{d}\mathds{Q}$ and next compare them to read an extended temperature $\mathpzc{T}$ as
\begin{equation}
\mathrm{d}\mathds{S}(\tau)= \frac{1}{\mathpzc{T}(\tau)}\mathrm{d}\mathds{Q}(\tau) + \mathrm{d}\Sigma(\tau).
\label{entro_prod-1}
\end{equation}

\begin{remark} \label{rem-t}
Note that Eq. (\ref{entro_prod-1}) defines both the extended temperature $\mathpzc{T}(\tau)$ and the generalized entropy production $\mathrm{d}\Sigma(\tau)$. Moreover, unlike the pseudo-temperature $T(\tau)$, $\mathpzc{T}(\tau)$ is by construction $\alpha_{S,B}$-independent because neither heat nor entropy depends on $\alpha_{S,B}$. In the following examples, we discuss both nonequilibrium temperatures $T(\tau)$ and $\mathpzc{T}(\tau)$ by comparing them with the equilibrium temperature $T$ (of the bath).
\end{remark}

\section{Examples}
\label{sec:examples}

Here we study in detail two examples in one of which \textit{thermalization} occurs, whereas the other one does not exhibit this feature.

\subsection{Example I: Thermalizing qubit}
\label{sec:ex1}

Consider a two-state system (e.g., a spin-$1/2$ particle or a two-level atom) interacting with a thermal environment, comprised of infinitely many modes at (initial) temperature $T=1/\beta$, through the Jaynes-Cummings total Hamiltonian $H=H_{0}+H^{(\lambda)}_{\mathrm{int}}$, where
\begin{align}
H_{0}&=\frac{1}{2}\omega_0 \sigma_z + \sum_{k=1}^{\infty}\omega_k \mathrm{a}^{\dag}_k \mathrm{a}_k,\\
H^{(\lambda)}_{\mathrm{int}}& = \lambda\sum_{k}(f^{*}_k \sigma_{+}\otimes\mathrm{a}_{k} + f_{k}\sigma_{-}\otimes\mathrm{a}^{\dag}_{k}).
\end{align}
Here $\sigma_{x}$, $\sigma_{y}$, and $\sigma_{z}=\mathrm{diag}(1,-1)$ are the Pauli operators, $\sigma_{\pm}=\sigma_{x}\pm i\sigma_{y}$, and $\mathrm{a}_{k}$ is the bosonic annihilation operator for mode $k$. Although this model is not exactly solvable, we can find the exact states of the system and bath up to any order in $\lambda$; see Appendix \ref{app:1} for details of $O(\lambda^{3})$ calculations. 

In the weak-coupling, long-time, $\omega$-continuum, Markovian limit (where $\lambda\to 0$ and $\tau\to\infty$ such that $\lambda^{2}\tau=\mathrm{const.}$ and $\sum_{k}\to\int_{0}^{\infty}\mathrm{d}\omega$), we can find the following Lindblad-type dynamical equation:
\begin{align}
&\frac{\mathrm{d}}{\mathrm{d}\tau} \varrho_{S}^{(\lambda)}(\tau) = -i\big[H_S+H_{\mathrm{LS}},\varrho^{(\lambda)}_{S}(\tau)\big] +\frac{\gamma}{2} \big(\overline{n}(\omega_0,\beta) + 1\big)\nonumber\\
&~~\times \Big(2\sigma_{-}\varrho^{(\lambda)}_S(\tau) \sigma_{+} -\{\sigma_{+}\sigma_{-},\varrho^{(\lambda)}_{S}(\tau)\} \Big) + \frac{\gamma}{2}\, \overline{n}(\omega_0,\beta)\nonumber\\
&~~\times\Big(2\sigma_{+}\varrho^{(\lambda)}_S(\tau) \sigma_{-} -\{\sigma_{-}\sigma_{+},\varrho^{(\lambda)}_{S}(\tau)\} \Big),
\label{lindblad-thermal-2}
\end{align}
where
\begin{align}
H_{\mathrm{LS}} &= 2\lambda^{2}\mathbb{P}\int_{0}^{\infty}\mathrm{d}\omega\,\frac{|f(\omega)|^{2}}{\omega_{0}-\omega}\left(2\,\overline{n}(\omega,\beta) +1\right)\,\sigma_{z}\\
&=:(1/2)\Omega(\lambda^{2},\omega_{0},\beta)\,\sigma_{z},
\end{align}
is the Lamb-shift Hamiltonian, $\mathbb{P}$ denotes the Cauchy principal value, $\beta$ is the inverse temperature of the bath, 
\begin{equation}
\overline{n}(\omega_0,\beta) = (\mathrm{e}^{\beta\omega_0}-1)^{-1}
\label{def:n}
\end{equation}
is the Planck distribution or the mean quanta number in a mode with frequency $\omega_{0}$, and 
\begin{align}
\gamma = 2\pi\lambda^{2}|f(\omega_{0})|^{2} 
\label{gamma-def}
\end{align}
is the spontaneous emission rate (see Appendix \ref{app:1}). This evolution agrees with the Markovian master equation derived in Ref.~\cite{Book:Carmichael}. The solution to Eq.~(\ref{lindblad-thermal-2}) is given by 
\begin{widetext}
\begin{equation}
\varrho^{(\lambda)}_{S}(\tau)= \frac{1}{2} 
    \begin{bmatrix}
    &1+ z(0)\mathrm{e}^{- \widetilde{\gamma}\tau}+ \tanh(\beta \omega_0 /2)\left( \mathrm{e}^{- \widetilde{\gamma}\tau} -1 \right)  & \big(x(0)- i y(0)\big)\, \mathrm{e}^{-i(\omega_0+\Omega) \tau- \widetilde{\gamma}\tau/2}  \\
    &\big(x(0) + i y(0)\big)\, \mathrm{e}^{i(\omega_0+\Omega) \tau- \widetilde{\gamma}\tau/2}   & 1- z(0)\mathrm{e}^{- \widetilde{\gamma}\tau}- \tanh(\beta \omega_0 /2)\left( \mathrm{e}^{- \widetilde{\gamma}\tau} -1 \right)
    \end{bmatrix},
\label{solution}
    \end{equation}
\end{widetext}
where $\widetilde{\gamma}= \gamma \coth(\beta \omega_0 /2) $. It is evident from this solution that system $S$ eventually thermalizes,
\begin{equation}
\lim_{\tau\to\infty}\varrho^{(\lambda)}_{S}(\tau) = \varrho_{S}^{\beta},
\end{equation}
where $\varrho_{S}^{\beta} = (1/Z_{\beta})\mathrm{e}^{-\beta\omega_{0}\sigma_{z}/2}$ is a thermal state in the Gibbs form, in which $Z_{\beta}=\mathrm{Tr}[\mathrm{e}^{-\beta\omega_{0}\sigma_{z}/2}]$ is the partition function.

We can explicitly compute $\mathrm{d}\mathds{S}^{(\lambda)}_S(\tau)$ as
\begin{align}
\mathrm{d}\mathds{S}^{(\lambda)}_S(\tau)=-\frac{1}{2}\log\frac{1+r^{(\lambda)}_{S}(\tau)}{1-r^{(\lambda)}_{S}(\tau)}\,\mathrm{d}r^{(\lambda)}_{S}(\tau),
\label{ds-S-1}
\end{align}
where $r^{(\lambda)}_{S}(\tau)$ is the norm of the Bloch vector $\mathbf{r}^{(\lambda)}_{S}=(x,y,z)$ associated with $\varrho^{(\lambda)}_{S}(\tau)$ as $\varrho_S = (1/2)(\openone + \mathbf{r}\cdot\bm{\sigma})$ (here $\bm{\sigma}=(\sigma_{x},\sigma_{y},\sigma_{z})$), and from Eq.~(\ref{lindblad-thermal-2}) we have
\begin{equation}
\mathrm{d}r_{S}(\tau) = \frac{\widetilde{\gamma}[x^2(\tau)+y^2(\tau)]-2\gamma z(\tau)-2\widetilde{\gamma}z^2(\tau)}{2\sqrt{x^2(\tau)+y^2(\tau)+z^2(\tau)}}\mathrm{d}\tau.
\end{equation}
Additionally, in the Markovian limit, the energy of this system is obtained as
\begin{equation}
\mathrm{d}\mathds{U}^{(\lambda)}_S(\tau)=  -\frac{\omega_0}{2}\gamma \mathrm{e}^{- \widetilde{\gamma}\tau}\big( \coth(\beta \omega_0 /2)z(0) + 1\big) \mathrm{d}\tau.
\end{equation}
As a result, 
\begin{equation}
\lim_{\tau\to\infty}\frac{1}{T^{(\lambda)}_{S}(\tau)} =\beta  \Big[ 1 - \frac{\big(x^2(0)+y^2(0)\big)\coth(\beta \omega_0 /2)}{2\big(z(0)+ \tanh(\beta \omega_0 /2)\big)} \Big].
\end{equation}
This pseudo-temperature behaves well, i.e., exhibits thermalization, if there is no initial coherence ($\varrho_{10}=0$, or equivalently, $x(0)=y(0)=0$). 

In the Markovian regime we consider the thermal bath always in equilibrium (namely, $\varrho_{B}(\tau)\approx\varrho_{B}^{\beta}$), and as a consequence the effective energy of $S$ reduces to (see Appendix \ref{app:1}, Eq.~(\ref{h-eff-s}))
\begin{equation}
\mathds{U}_S(\tau) = \mathrm{Tr}\left[ \varrho_S(\tau) H_S \right],
\label{u-s}
\end{equation} 
and the heat flux reads
\begin{equation}\label{QS}
\mathrm{d}\mathds{Q}_S(\tau)= \mathrm{Tr}\left[ \mathrm{d}\varrho_S(\tau) H_S \right]= \frac{\omega_0}{2}\mathrm{d}z(\tau).
\end{equation}
Comparing Eqs. \eqref{ds-S-1} and \eqref{QS} yields
\begin{align}
\frac{1}{\mathpzc{T}_{S}(\tau)}&= -\frac{1}{\omega_0}\frac{ z(\tau) }{r_{S}(\tau)}\log\frac{1+r_{S}(\tau)}{1-r_{S}(\tau)},
\label{temperature}\\
\mathrm{d}\Sigma_{S}(\tau) &= -\frac{1}{2} \frac{\mathrm{d}r_{S}(\tau)}{r_{S}(\tau)}\log\frac{1+r_{S}(\tau)}{1-r_{S}(\tau)},
\end{align}
which are both $\alpha_{S,B}$-independent. By substituting the Bloch vector components of the Gibbs state $\varrho^{\beta}_{S}$, $(x=0,y=0,z=-\tanh(\beta \omega_{0} /2))$, in Eq. \eqref{temperature}, we also see that $\lim_{\tau\to\infty}\mathpzc{T}_{S}(\tau)=T$, which gives the expected equilibrium temperature.

For the bath thermodynamics, after some algebra (see Appendix \ref{app:1}) we find that when $\tau\to \infty$ (up to $O(\lambda^{3})$)
\begin{align}
\mathrm{d}\mathds{Q}^{(\lambda)}_B(\tau)=& 4\omega_{0}\gamma\Big(  \big[\big( \overline{n}(\omega_{0},\beta)+1\big)\varrho_{00} -\overline{n}(\omega_{0},\beta) \varrho_{11} \big]\nonumber\\
& - |\varrho_{10}|^{2} \Big)\mathrm{d}\tau,\\
\mathrm{d}\mathds{U}^{(\lambda)}_{B}(\tau)=& 4\gamma\omega_{0}\left[\big( \overline{n}(\omega_{0},\beta)+1\big)\varrho_{00} -\overline{n}(\omega_{0},\beta) \varrho_{11} \right]\mathrm{d}\tau,\label{ex:duB}\\
\mathrm{d}\mathds{S}^{(\lambda)}_B(\tau)=& 4\beta\gamma\omega_{0}\big[\big( \overline{n}(\omega_{0},\beta)+1\big)\varrho_{00} -\overline{n}(\omega_{0},\beta) \varrho_{11} \nonumber\\
&-|\varrho_{10}|^{2}\big]\mathrm{d}\tau,
\label{ex:dsB}
\end{align}
whence 
\begin{align}
\lim_{\tau\to\infty}T_{B}^{(\lambda)}(\tau) &= \frac{1}{\beta} \Big[ 1+\frac{ |\varrho_{10}|^{2} }{ \overline{n}(\omega_{0},\beta) (\varrho_{00} - \varrho_{11})+\varrho_{00} - |\varrho_{10}|^{2} }\Big],\label{tb-limit}\\
\mathrm{d}\mathds{S}^{(\lambda)}_B(\tau)&= \beta \,\mathrm{d}\mathds{Q}^{(\lambda)}_B(\tau). \label{ex-tp}
\end{align}
Note that the limit (\ref{tb-limit}) is independent of $\alpha_{B}$ but it depends on the initial state of the system. However, if $\varrho_S(0)$ does not have any coherence, i.e., $\varrho_{10}=0$, one retrieves the expected value $1/\beta$ for the pseudo-temperature $T_{B}^{(\lambda)}$. But regardless of the initial state of the system, from Eq.~(\ref{ex-tp}), we see that the extended temperature behaves as expected, $\lim_{\tau\to\infty}\mathpzc{T}^{(\lambda)}_{B}(\tau)=T$. Besides, the internal entropy production of the bath up to $O(\lambda^{3})$ vanishes,
\begin{equation}
\mathrm{d}\Sigma^{(\lambda)}_{B}=0.
\end{equation}

\begin{remark}
Following the discussion in Remark \ref{rem:2}, the reason for the difference between the pseudo-temperature and the standard thermodynamic temperature lies in the definition of the former. The entropy of the qubit in this example can be computed using its eigenvalues, which in general depend on the $(x,y,z)$ components of the Bloch vector. From Eq.~(\ref{u-s}), we can identify the $z$ component with $\mathds{U}$. Thus we can say that $\mathds{S}$ is a function of $(x,y,\mathds{U})$, and we can compute the partial derivative of $\mathds{S}$ with respect to $\mathds{U}$ (while keeping $x$ and $y$ fixed),
\begin{align}
\left(\frac{\partial \mathds{S}(x,y,\mathds{U})}{\partial \mathds{U}} \right)_{x,y} =&-\frac{1}{2}\frac{\mathds{U}}{ \sqrt{x^2+y^2+\mathds{U}^2} }\nonumber\\
&\times\log\frac{1+\sqrt{x^2+y^2+\mathds{U}^2}}{1-\sqrt{x^2+y^2+\mathds{U}^2}}.
\end{align}
If we now consider $x(\tau)$, $y(\tau)$, and $\mathds{U}(\tau)$ evolving according to the dissipative thermalizing dynamics \eqref{solution}, we obtain 
\begin{equation}
\lim_{\tau \to \infty}\left(\frac{\partial \mathds{S}(x,y,\mathds{U})}{\partial \mathds{U}}\right)_{x,y}=\beta,
\end{equation}
which agrees with the standard definition of the equilibrium temperature. Rather, the inverse pseudo-temperature $1/T^{(\lambda)}_{S}(\tau)$ reads as
\begin{equation}
\frac{\mathrm{d} \mathds{S}\big(x(\tau),y(\tau),z(\tau)\big)/\mathrm{d}\tau}{\mathrm{d} \mathds{U}/\mathrm{d}\tau},
\end{equation}
which corresponds to inverting the function $\mathds{U}(\tau)$, finding $\tau(\mathds{U})$, and computing the total derivative with respect to $\mathds{U}$,
\begin{align}
\frac{\mathrm{d} \mathds{S}}{\mathrm{d} \mathds{U}}\big(x(\mathds{U}),y(\mathds{U}\big),\mathds{U})= &-\frac{1}{2}\frac{\mathds{U} + x(\mathds{U})\frac{\mathrm{d}x}{\mathrm{d}\mathds{U}}+y(\mathds{U})\frac{\mathrm{d}y}{\mathrm{d}\mathds{U}}}{ \sqrt{x^2(\mathds{U})+y^2(\mathds{U})+\mathds{U}^2} } \nonumber\\
&\times\log\frac{1+\sqrt{x^2(\mathds{U})+y^2(\mathds{U})+\mathds{U}^2}}{1-\sqrt{x^2(\mathds{U})+y^2(\mathds{U})+\mathds{U}^2}}.
\end{align}
In the $\tau\to\infty$ limit (or $\mathds{U} \to\mathds{U}_{\mathrm{thermal}}$) this derivative is different from $\beta$ because in general
\begin{equation}
x(\mathds{U})\frac{\mathrm{d}x}{\mathrm{d}\mathds{U}}+y(\mathds{U})\frac{\mathrm{d}y}{\mathrm{d}\mathds{U}}=\frac{x^2(0)+y^2(0)}{2\big(z(0)+ \tanh(\beta \omega_0 /2)\big)} \neq 0.
\end{equation}
The two derivatives coincide only if $x$ and $y$ are fixed during the dynamics, which is the case of vanishing initial coherence.
\end{remark}

\subsection{Example II: Dephasing qubit}
\label{sec:ex2}

We apply the previous considerations to the exactly solvable model of a qubit in interaction with a thermal bosonic bath \cite{Palma}. The total Hamiltonian is given by $H_{\mathrm{tot}}=H_0+H^{(\lambda)}_{\mathrm{int}}$ with
\begin{equation*}
H_0=\frac{\omega_{0}}{2}\sigma_z+\sum_{k=1}^{\infty}\omega_k \mathrm{a}^\dag_k \mathrm{a}_k\ ,\quad
H^{(\lambda)}_{\mathrm{int}}=\lambda\sigma_z\otimes\big(\mathrm{a}(f)+\mathrm{a}^\dag(f)\big),
\end{equation*}
where $\mathrm{a}_k$ is the bosonic annihilation operator of mode $k$, satisfying the commutation relations $[\mathrm{a}_k,\mathrm{a}^\dag_l]=\delta_{kl}$, and 
\begin{equation}
\mathrm{a}(f)=\sum_k f_k^*\mathrm{a}_k,
\label{notation:af}
\end{equation}
with complex quantities $f_k$ forming a square-summable vector $f=\{f_k\}\in \mathpzc{L}^2(-\infty,+\infty)$ such that
\begin{equation}
\label{CCR}
\left[\mathrm{a}(f),\mathrm{a}^\dag(g)\right]=\langle f\vert g\rangle.
\end{equation}
In the interaction picture, the Hamiltonian becomes
\begin{equation}
\widetilde{H}^{(\lambda)}_{\mathrm{int}}(\tau)=U^\dag_0(\tau)H^{(\lambda)}_{\mathrm{int}}U_0(\tau)=\lambda\sigma_z\otimes\big(\mathrm{a}(f_{\tau})+\mathrm{a}^\dag(f_{\tau})\big),
\end{equation}
where $U_0(\tau)=\mathrm{e}^{-iH_0\tau}$ and $f_{\tau}$ is the vector with components $f_k^*\mathrm{e}^{-i\omega_k \tau}$. The time-ordered exponentiation of $\widetilde{H}^{(\lambda)}_\mathrm{int}(\tau)$ yields
\begin{align*}
\widetilde{U}_\lambda(\tau)&=\mathbb{T}\mathrm{e}^{-i\lambda\sigma_z\otimes \int_{0}^{\tau}\mathrm{d}s\,\left(\mathrm{a}(f_s)+\mathrm{a}^\dag(f_s)\right)}\\
&=\mathrm{e}^{-i\lambda^2\varphi(\tau)}\mathrm{e}^{-i\lambda\sigma_z\otimes \int_{0}^{\tau}\mathrm{d}s\,\left(\mathrm{a}(f_s)+\mathrm{a}^\dag(f_s)\right)},
\end{align*}
where the pure phase factor
\begin{equation}
\varphi(\tau)=\sum_k\frac{|f_k|^2}{\omega_k^2}\big(\omega_k\tau-\sin(\omega_k \tau)\big)
\end{equation}
does not contribute to the evolution
\begin{equation}
\label{PSE}
\varrho^{(\lambda)}_{SB}(\tau)=U_0(\tau)\widetilde{U}_\lambda(\tau)\varrho_{SB}(0) \widetilde{U}_\lambda^\dag(\tau)U^\dag_0(\tau).
\end{equation}
Let us assume that the initial state of the total system is factorized and given by $\varrho_{SB}(0)=\varrho_S(0)\otimes \varrho_B^\beta$, where 
\begin{equation}
\label{Sst}
\varrho_S(0)=\sum_{\ell,\ell'=0}^1\varrho_{\ell\ell'}\vert\ell\rangle\langle\ell'\vert\ ,\qquad  \sigma_z\ket{\ell}=(-)^\ell\ket{\ell},
\end{equation} 
is the initial state of the qubit, and $\varrho_{B}^{\beta}$ is the Gibbs (thermal) state of the bosonic bath with the inverse temperature $\beta$,
\begin{equation}
\label{thermst}
\varrho_B^\beta=\mathrm{e}^{-\beta\sum_k\omega_k\mathrm{a}^\dag_k\rm{a}_k}/Z_\beta,
\end{equation}
and $Z_{\beta}=\mathrm{Tr}[\mathrm{e}^{-\beta\sum_k\omega_k \mathrm{a}^\dag_k\mathrm{a}_k}]$ is the associated partition function. One can see that  
\begin{equation}
\varrho^{(\lambda)}_{SB}(\tau)=
\sum_{\ell,\ell'=0}^1\varrho_{\ell\ell'}\mathrm{e}^{-i\omega_{0}\zeta_{\ell \ell'}\tau/2}\ketbra{\ell}{\ell'}\otimes  D_{\ell}(g_\tau)\,\varrho_{B}^{\beta}\, D^{\dag}_{\ell'}(g_\tau),
\label{total evol}
\end{equation}
where $\zeta_{\ell\ell'}=(-)^{\ell}-(-)^{\ell'}$, $g_\tau$ is the vector with components
\begin{equation}
g^{*}_k(\tau)=f_k^*(\mathrm{e}^{-i\omega_k\tau}-1)/\omega_k,
\label{notation:gk}
\end{equation}
and $D_\alpha(g_\tau)$ is the displacement operator
\begin{equation}
D_{\ell}(g_\tau)= \mathrm{e}^{(-)^\ell\lambda[\mathrm{a}^\dag(g_\tau)-\mathrm{a}(g_\tau)]},
\end{equation}
whose action can be derived from Eq.~\eqref{CCR} as
\begin{align}
\label{displ}
D_{\ell}(g_\tau)\,\mathrm{a}_k\,D^\dag_{\ell}(g_\tau)&=\mathrm{a}_k-(-)^\ell \lambda\, g_k(\tau)\\
&=:A_k(\ell,\lambda,\tau). \label{newA}
\end{align}
From here the reduced density matrices of the two subsystems read as
\begin{align}
\varrho^{(\lambda)}_S (\tau)=&\varrho_{00}\ketbra{0}{0}+\varrho_{11}\ketbra{1}{1} + \mathrm{e}^{-8\lambda^2\Gamma(\tau)}\nonumber \\
\label{Sevol}
&\times\left(\varrho_{10}\mathrm{e}^{i\omega_{0}\tau}\ketbra{1}{0}+\varrho_{01}\mathrm{e}^{-i\omega_{0}\tau}\ketbra{0}{1}\right),\\
\varrho^{(\lambda)}_{B}(\tau)=&\sum_{\ell=0}^{1}\varrho_{\ell\ell}D_\ell(g_\tau)\,\varrho_B^\beta \,D_\ell^\dag(g_\tau)\nonumber\\
=&\sum_{\ell=0}^{1}\frac{\varrho_{\ell\ell}}{Z_\beta}\mathrm{e}^{-\beta\sum_k\omega_k
A^\dag_k(\ell,\lambda,\tau)\,A_k(\ell,\lambda,\tau)},\label{Bevol}
\end{align}
where $\mathrm{Tr}\big[D_{\ell}(g_\tau)\,\varrho_B^\beta\, D^{\dag}_{\ell}(g_\tau)\big]= \mathrm{e}^{-8\lambda^2\Gamma(\tau)}$ for $\alpha\neq\delta$, with
\begin{equation}
\label{Gamma}
\Gamma(\tau)=\sum_k\frac{|f_k|^2}{\omega_k^2}\coth(\beta\omega_k/2)\sin^2(\omega_k\tau/2).
\end{equation}

Further, denoting the qubit polarization at time $\tau=0$ by $\average{\sigma_z}_S$, the effective qubit Hamiltonian takes the form
\begin{equation}
H^{(\mathrm{eff})}_S(\tau)= \big(\frac{\omega_{0}}{2} -4\lambda^2\langle\sigma_z\rangle_S\,\Delta(\tau)\big)\sigma_z +4 \lambda^2 \alpha_S \average{\sigma_z}^2_S \Delta(\tau),
\label{q eff}
\end{equation}
where the explicit time dependence is provided by 
\begin{align}
\Delta(\tau)&=-\frac{1}{4\lambda\langle\sigma_z\rangle_S}\,\mathrm{Tr}\left[\varrho^{(\lambda)}_B(\tau)\big(\mathrm{a}(f)+\mathrm{a}^\dag(f)\big)\right]\nonumber \\
&=\sum_k\frac{|f_k|^2}{\omega_k}\sin^2(\omega_k\tau/2).
\label{Delta}
\end{align}
Similarly, the bath effective Hamiltonian reads as 
\begin{align}
H^{(\mathrm{eff})}_B(\tau)=& \sum_k\omega_k \mathrm{a}^\dag_k \mathrm{a}_k+\lambda\average{\sigma_z}_S\left(\mathrm{a}(f)+\mathrm{a}^\dag(f)\right)\nonumber \\
&+4\lambda^2\alpha_B\average{\sigma_z}^2_S \Delta(\tau),
\label{F eff}
\end{align}
where the time-dependent appears only in the scalar term. From the above relations, the exchanged works between system $S$ and bath $B$ are calculated by using Eq.~\eqref{work}, 
\begin{equation}
\label{sbW}
\mathrm{d}\mathds{W}^{(\lambda)}_{B}(\tau)=4\lambda^2\alpha_B \average{\sigma_z}^2_S \mathrm{d}{\Delta}(\tau)=-\mathrm{d}\mathds{W}^{(\lambda)}_S(\tau),
\end{equation}
where the last equality verifies Eq.~\eqref{workSB2}. In addition, using Eq.~\eqref{heatSB1} and the fact that
\begin{align}
\big[H_S^{(\mathrm{eff})}(\tau),H_{\mathrm{int}}\big]&=0, \nonumber \\
\big[H_B^{(\mathrm{eff})}(\tau),H_{\mathrm{int}}\big]&=\lambda\sigma_z\otimes\sum_k\omega_k(
f_k\mathrm{a}^\dag_k-f^*_k\mathrm{a}_k), \nonumber
\end{align}
the infinitesimal heat exchanges are given by
\begin{align}
\label{sQ}
\mathrm{d}\mathds{Q}^{(\lambda)}_S(\tau)&=0,\\
\label{bQ}
\mathrm{d}\mathds{Q}^{(\lambda)}_B(\tau)&=4\lambda^2\big(1-\average{\sigma_z}_S^2\big)\,\mathrm{d}\Delta(\tau).
\end{align}
The binding energy also becomes
\begin{equation}
\mathds{U}^{(\lambda)}_\chi(\tau)=-4\lambda^2\big(1-\average{\sigma_z}_S^2\big)\,\Delta(\tau),
\end{equation}
whence $\mathrm{d}\mathds{Q}^{(\lambda)}_B(\tau)=-\mathrm{d}\mathds{U}^{(\lambda)}_\chi(\tau)$, in agreement with Eq.~\eqref{heatSB2}.

Equation (\ref{sQ}) is physically expected because, with our specific system Hamiltonian ($H_S\propto\sigma_z$) and the interaction Hamiltonian ($H_{\mathrm{int}}\propto \sigma_{z}\otimes(\mathrm{a}+\mathrm{a}^{\dag})$), we have $[H_{S},H_{\mathrm{int}}]=0$. That is, this interaction with the environment cannot excite or change the \textit{populations} of $\varrho_S(0)$; $\varrho_{00}(\tau)=\varrho_{00}$ [Eq.~(\ref{Sevol})]. Thus according to the definition of the heat exchange,  we should have $\mathrm{d}\mathds{Q}^{(\lambda)}_S(\tau)=\mathrm{Tr}[\mathrm{d}\varrho_{S}^{(\lambda)}(\tau)H_{S}^{\mathrm{(eff)}}(\tau)]=\sum_{\ell=0}^{1}\mathrm{d}\varrho_{\ell\ell}(\tau)q(\tau)\sigma_{z,~\ell\ell}=0$, where we have used the fact that $H_{S}^{\mathrm{(eff)}}(\tau)=q(\tau)\sigma_{z}$ (for some appropriate $q$ read through Eq.~(\ref{q eff})).

Furthermore, using Eqs.~\eqref{sbW} and \eqref{bQ}, and the fact that $\alpha_S+\alpha_B=1$, it turns out that, unlike the infinitesimal heat exchanges, the infinitesimal variation of the internal energy of $B$ depends on $\alpha_S$,
\begin{align}
\mathrm{d}\mathds{U}^{(\lambda)}_B(\tau)= 4\lambda^2\big(1-\alpha_S\average{\sigma_z}_S^2\big)\,\mathrm{d}\Delta(\tau).
\label{uB}
\end{align}
One expects the final pseudo-temperature of $T^{(\lambda)}_{B}(\infty)$---as defined by Eq.~\eqref{T-def}---to tend to the (initial) bath temperature $T=1/\beta$ in the limit $\lambda\to0$ of vanishing coupling between $S$ and $B$. Indeed, if $\lambda=0$, the thermal state \eqref{thermst} is time-invariant. Since $\mathrm{Tr}[\mathrm{d}\varrho_B^{(\lambda)}(\tau)]=0$, the infinitesimal variation of the von Neumann entropy of $B$ is given by
\begin{equation}
\label{vNeB}
\mathrm{d}\mathds{S}^{(\lambda)}_B(\tau)=-\mathrm{Tr}\left[\mathrm{d}\varrho_B^{(\lambda)}(\tau)\log\varrho_B^{(\lambda)}(\tau)\right].
\end{equation}
By expanding Eq.~\eqref{Bevol} up to $O(\lambda^3)$ one obtains (see Appendix \ref{app:2})
\begin{align}
\mathrm{d}\mathds{S}_{B}^{(\lambda)}(\tau) = 4\beta\lambda^2\big(1-\langle\sigma_z\rangle_S^2\big)\,\mathrm{d}\Delta(\tau).\label{dsb}
\end{align}
Now if we use Eqs.~(\ref{sbW}) and (\ref{bQ}), together with the definition of the pseudo-temperature (\ref{T-def}), we obtain
\begin{align}
\lim_{\tau\to\infty}T_{B}^{(\lambda)}(\tau)=\frac{(1-\alpha_S \langle \sigma_z\rangle^2_S)}{(1-\langle \sigma_z\rangle^2_S)}\,T.
\end{align}
It is evident from this expression that in order to make the pseudo-temperature $T^{(\lambda)}_{B}$ to be equal to $T$ (in the weak-coupling limit) we need to set $\alpha_S=1$. 

Additionally, we note that by comparing Eqs.~(\ref{bQ}) and (\ref{dsb}), these quantities are related as
\begin{equation}
\mathrm{d}\mathds{S}^{(\lambda)}_B(\tau)= \beta\, \mathrm{d}\mathds{Q}^{(\lambda)}_B(\tau).
\end{equation} 
Hence, we have $\mathpzc{T}_{B}(\tau)=T$ and the inverse temperature $\beta=1/T$ of the bath shows up as the prefactor of the heat flux, as expected in the standard equilibrium thermodynamics (Eq.~(\ref{entro_prod})). Thus up to $O(\lambda^3)$ the internal entropy production in the bath vanishes,
\begin{equation}
\mathrm{d}\Sigma_{B}^{(\lambda)}(\tau)=0.
\end{equation}
This is consistent with the classical picture where the bath always exchanges heat quasistatically---see the discussion at the end of Sec.~\ref{sec:2nd-law}. 

\begin{remark} 
We have verified in two different models that the internal entropy production in a thermal bath vanishes in the weak-coupling limit up to the leading order in $\lambda$. This seems to be a general result and is consistent with our expectation from standard, equilibrium thermodynamics.
\end{remark}

Now we consider the pseudo-temperature $T^{(\lambda)}_S(\tau)$. We first note that, from  Eqs.~\eqref{sbW} and \eqref{sQ} and after setting $\alpha_S=1$, we have 
\begin{equation}
\mathrm{d}\mathds{Q}^{(\lambda)}_{S}(\tau)= \mathrm{d}\mathds{W}^{(\lambda)}_{S}(\tau)=0,
\end{equation}
and thus
\begin{equation}
\label{sW}
\mathrm{d}\mathds{U}^{(\lambda)}_S(\tau)=0.
\end{equation}
That is, despite interacting with bath $B$, system $S$ does not exchange any heat or work (and thus internal energy) with $B$. Hence intuitively we should not expect that its temperature $T_{S}^{(\lambda)}$ to change; it should remain constant. This is explicitly seen by calculating 
\begin{equation}
T_{S}^{(\lambda)}(\tau)=\frac{\mathrm{d}\mathds{U}_{S}^{(\lambda)}(\tau)}{\mathrm{d}\mathds{S}_{S}^{(\lambda)}(\tau)}=0.
\end{equation}
Note that if the system were initially prepared, e.g., in a thermal state with temperature $T^{(0)}_{S}\neq0$, in principle its temperature should not change because this system does not thermalize [Eq.~(\ref{Sevol})]. This fact is captured by our pseudo-temperature as $T^{(\lambda)}_{S}=0$. However, we note that $T(\tau)$ is defined by the given \textit{dynamics} of $S$ and cannot therefore be related to an initial (dynamics-independent) temperature such as an equilibrium temperature assigned to the preparation of the state.

Having calculated the heat and work exchanges by the system, it is also important to see how entropy of the system behaves. Using Eq.~\eqref{Sevol}, the entropy of $S$ can be explicitly calculated from the eigenvalues $\big(1\pm r_S^{(\lambda)}(\tau)\big)/2$ of $\varrho^{(\lambda)}_S(\tau)$, where
\begin{equation}
r_S^{(\lambda)}(\tau)=\sqrt{1-4\left(\varrho_{00}\varrho_{11}-\mathrm{e}^{-16\lambda^2\Gamma(\tau)}|\varrho_{01}|^2\right)},
\end{equation}
as well as its infinitesimal variation
\begin{align}
\label{dvNeS0}
\mathrm{d}\mathds{S}_S^{(\lambda)}(\tau)&=-\frac{1}{2}\log\frac{1+r^{(\lambda)}_S(\tau)}{1-r_S^{(\lambda)}(\tau)}\,\mathrm{d}r_S^{(\lambda)}(\tau)=\lambda^{2}b^{(\lambda)}(\tau)\mathrm{d}\Gamma(\tau),
\end{align}
where
\begin{align}
b^{(\lambda)}(\tau)=\frac{16|\varrho_{01}|^2}{r_S^{(\lambda)}(\tau)}\mathrm{e}^{-16\lambda^2\Gamma(\tau)}\log\frac{1+r^{(\lambda)}_S(\tau)}{1-r_S^{(\lambda)}(\tau)}.
\label{dvNeS}
\end{align}
Note that the quantity $b^{(\lambda)}(\tau)$ is nonnegative and has a well-defined time-independent limit,
\begin{equation}
b^{(0)}=\frac{16|\varrho_{01}|^2}{r^{(0)}_S(0)}\log\frac{1+r^{(0)}_S(0)}{1-r^{(0)}_S(0)},
\end{equation}
when $\lambda\to0$. In order to study the time-derivatives $\mathrm{d}\Delta(\tau)/\mathrm{d}\tau$ and $\mathrm{d}\Gamma(\tau)/\mathrm{d}\tau$, we consider an \textit{infinite} thermal bath with a \textit{continuum} $\omega$ and a regularized Ohmic spectral density given by $f_k\simeq\sqrt{\omega}\mathrm{e}^{-\omega\epsilon/2}$ (in which $\epsilon\geqslant  0$). Thus we substitute the discrete sums in Eqs.~\eqref{Gamma} and \eqref{Delta} with the following integrals:
\begin{eqnarray}
\label{Delta1}
\hskip-.5cm
\Delta(\tau)&=&\int_0^{\infty}\mathrm{d}\omega\,\sin^2(\omega\tau/2)~\mathrm{e}^{-\epsilon\omega}=\frac{\tau^2}{2\epsilon(\epsilon^2+\tau^2)},\\
\label{Gamma1}
\hskip-.5cm
\Gamma(\tau)&=&\int_0^{\infty}\mathrm{d}\omega
\frac{1}{\omega} \coth(\beta\omega/2)\sin^2(\omega\tau/2)~\mathrm{e}^{-\epsilon\omega}. 
\end{eqnarray}
Hence, $\mathrm{d}\Delta(\tau)/\mathrm{d}\tau=\epsilon\,\tau/(\tau^2+\epsilon^2)^2$ as well as $\mathrm{d}\Gamma(\tau)/\mathrm{d}\tau\geqslant 0$, as one can check by changing the variable $\omega\tau=\widetilde{\omega}$ and taking explicitly the derivative with respect to $\tau$. As a result, we see that $\mathrm{d}\mathds{S}^{(\lambda)}_{S}(\tau)\geqslant 0$. Furthermore, as a consequence of Eq.~\eqref{sQ}, in this regime, the internal entropy production relation \eqref{entbal} reduces to
\begin{equation}
\mathrm{d}\Sigma_S(\tau)=\mathrm{d} \mathds{S}^{(\lambda)}_S(\tau)\geqslant 0.
\end{equation}
That is, the whole entropy change in the system is entirely due to the internal entropy production, whence the extended temperature $\mathpzc{T}_{S}(\tau)$ remains undefined because of Remark \ref{rem-t}.

\begin{remark}
It is an appealing feature of this model that the qubit does not exchange any energy with its environment ($\mathrm{d}\mathds{U}_{S}^{(\lambda)}(\tau)=0$), whilst its (internal) entropy may change ($\mathrm{d}\mathds{S}_{S}^{(\lambda)}(\tau)\neq0$) completely because of its (varying) \textit{correlations} with the environment ($\mathrm{d}\mathds{S}_{\chi}(\tau)\neq0$).
\end{remark}

It also may also be interesting to investigate the behavior of the various thermodynamic quantities in the Markovian regime for system $S$. This is determined by the condition $\beta\ll\tau$ over the long timescale $1/\lambda^2$ when $\lambda\to0$. Under these conditions and after removal of the regularization parameter $\epsilon$, one obtains $\Gamma(\tau)\simeq \pi \tau/(2\beta)$. Thus, the dynamics of system $S$ [Eq.~(\ref{Sevol})] reads as
\begin{align}
\varrho^{(\lambda)}_S (\tau)=&\varrho_{00}\ketbra{0}{0}+\varrho_{11}\ketbra{1}{1} +\mathrm{e}^{-\gamma\tau}(\varrho_{10}\mathrm{e}^{i\omega_{0}\tau}\ketbra{1}{0}\nonumber\\
&+\varrho_{01}\mathrm{e}^{-i\omega_{0}\tau}\ketbra{0}{1}),
\label{sol-ex-2}
\end{align}
in which $\gamma=4\pi\lambda^2/\beta$. This state solves the Lindblad-type master equation
\begin{equation}
\label{LindspinBose}
\frac{\mathrm{d}\varrho^{(\lambda)}_S(\tau)}{\mathrm{d}\tau}=-i\Big[\frac{1}{2}\omega_{0}\sigma_z,\varrho^{(\lambda)}_S(\tau)\Big]+\frac{\gamma}{2}\big(\sigma_z\varrho^{(\lambda)}_S(\tau)\sigma_z-\varrho^{(\lambda)}_S(\tau)\big).
\end{equation}
Note that this dynamics, similarly to dynamics generated by a Lindblad equation, has a fixed point as $\lim_{\tau\to\infty}(1/\tau)\int_{0}^{\tau}\mathrm{d}s\,\varrho^{(\lambda)}_{S}(s)= \varrho_{00}\ketbra{0}{0}+\varrho_{11}\ketbra{1}{1}$. Thus if we start with the system initially with no coherence (i.e., vanishing off-diagonal elements, $\varrho_{10}=0$), it will not evolve in time, and because of Eq.~(\ref{dvNeS}) the bath will not experience any entropy change either; $\mathds{S}_{B}^{(\lambda)}(\tau)=\mathrm{const}$.

\section{Summary and outlook} 
\label{sec:summary}

This paper highlights the role of correlations in the nonequilibrium thermodynamic behavior of generic bipartite interacting quantum systems. In this formulation, interesting relations emerge between correlations, on the one hand, and heat, work exchanges, as well as possible definitions of nonequilibrium temperatures of each subsystem, on the other hand. These relations may enable the extraction of desired thermodynamic properties by partially controlling or manipulating the underlying dynamics of the system. A notion of binding energy has been introduced which only depends on the interaction Hamiltonian and correlations of the total system state, whose variation has been shown to be only of the heat type. In addition, this energy has been shown not to be locally accessible by the subsystems, but it provides a heat transmission channel between the parties. In this sense, correlations act as a resource or storage for heat. We have also defined two notions of nonequilibrium temperatures for the subsystems and discussed their relevance in the thermodynamic equilibrium. We also associated a nonequilibrium temperatures with correlations. This temperature may enable one to obtain conditions such that the two subsystems have same nonequilibrium temperatures, which are generically different exactly because of correlations. These results have been illustrated in detail through two examples: a qubit in interaction with a thermalizing bath and a qubit interacting with a dephasing environment.

Our methodology may provide techniques and tools for employing quantum resources, such as manybody correlations and memory, to engineer thermodynamic processes, for example, to build efficient quantum heat engines, or shed light on our understanding of the role of correlations in biological processes in relation to, e.g., the efficiency of photosynthetic light-harvesting complexes \cite{Plenio}.

\textit{Acknowledgements.}---Hospitality of the Abdus Salam International Center for Theoretical Physics (ICTP) and the Institute for Research in Fundamental Sciences (IPM) are acknowledged, respectively, by S.A. and S.M., where parts of this work were completed. A.T.R. acknowledges financial support by Sharif University of Technology's Office of Vice President for Research, the Iran Science Elites Federation, and the IPM.  

\appendix
\section{Details of example I}
\label{app:1}

\subsection{State of the system}

Here we obtain the exact state of the total system up to the second order in the interaction coupling $\lambda$. After calculating the interaction-picture Hamiltonian $\widetilde{H}_{\mathrm{int}}^{(\lambda)}(\tau)=U^{\dag}_{0}(\tau) H_{\mathrm{int}}^{(\lambda)} U_{0}(\tau)$ and the corresponding evolution operator $\widetilde{U}_{\lambda}(\tau)=\mathbb{T}\mathrm{e}^{-i\int_{0}^{\tau}\mathrm{d}s\,\widetilde{H}_{\mathrm{int}}^{(\lambda)}(s)}$, one can read the state of the combined system from 
\begin{equation}
\varrho^{(\lambda)}_{SB}(\tau)=U_0(\tau)\widetilde{U}_\lambda(\tau)\varrho_{SB}(0) \widetilde{U}_\lambda^\dag(\tau)U^\dag_0(\tau)
\label{g-dynamics}
\end{equation}
as
\begin{widetext}
\begin{align}
\varrho_{S}^{(\lambda)}(\tau) =&\varrho_{S}^{(0)}(\tau) + \lambda^2\Big\{ \sigma_{+}\varrho_{S}^{(0)}(\tau) \sigma_{-} \sum_{k}|f_{k}|^{2}\,|\eta(\omega_{0},\omega_k,\tau)|^{2} \,\overline{n}(\omega_{k},\beta) +\nonumber\\
& + \sigma_{-} \varrho_{S}^{(0)}(\tau) \sigma_{+} \sum_{k}|f_{k}|^{2} \,|\eta(\omega_{0},\omega_{k},\tau)|^{2}\big( \overline{n}(\omega_{k},\beta)+1\big) \nonumber\\
&  - \sum_{k}|f_k|^{2}\Big(  \xi^{*}(\omega_0,\omega_{k},\omega_{k},\tau)\,\varrho_{S}^{(0)}(\tau)\sigma_{+}\sigma_{-} + \xi(\omega_0,\omega_{k},\omega_{k},\tau)\,\sigma_{+}\sigma_{-}\,\varrho_{S}^{(0)}(\tau) \Big) \big(\overline{n}(\omega_{k},\beta)+1\big) \nonumber\\
&  - \sum_{k}|f_k|^{2}\Big(  \xi(\omega_0,\omega_{k},\omega_{k},\tau)\,\varrho_{S}^{(0)}(\tau)\sigma_{-}\sigma_{+} + \xi^{*}(\omega_0,\omega_{k},\omega_{k},\tau)\,\sigma_{-}\sigma_{+}\,\varrho_{S}^{(0)}(\tau) \Big) \overline{n}(\omega_{k},\beta) \Big\} +O(\lambda^3),\label{rhos-2}
\end{align}
and similarly for bath $B$,
\begin{align}
\varrho_{B}^{(\lambda)}(\tau) =& \varrho_{B}^{\beta} +i\lambda\Big(\mathrm{Tr}[\varrho_{S}(0)\sigma_{+}]\,\mathrm{e}^{i\omega_{0}\tau}\sum_{k}f^{*}_{k}\,\eta^{*}(\omega_{0},\omega_{k},\tau)\,[\varrho_{B}^{\beta},\mathrm{a}_{k}] + \mathrm{Tr}[\varrho_{S}(0)\sigma_{-}]\,\mathrm{e}^{-i\omega_{0}\tau}\sum_{k}f_{k}\,\eta(\omega_{0},\omega_{k},\tau)\,[\varrho_{B}^{\beta},\mathrm{a}^{\dag}_{k}]  \Big)\nonumber\\
&+\lambda^{2}\Big( \mathrm{Tr}[\varrho_{S}(0)\sigma_{-}\sigma_{+}]\sum_{kk'}\Big\{f_{k}^{*}f_{k'}\,\eta^{*}(\omega_{0},\omega_{k},\tau)\eta(\omega_{0},\omega_{k'},\tau)\,\mathrm{a}_{k}\,\varrho_{B}^{\beta}\,\mathrm{a}^{\dag}_{k'}\nonumber\\
& -f_{k'}f_{k}^{*} \xi^{*}(\omega_{0},\omega_{k'},\omega_{k},\tau)\,\mathrm{e}^{i\tau(\omega_{k}-\omega_{k'})}\,\mathrm{a}^{\dag}_{k'}\mathrm{a}_{k}\,\varrho_{B}^{\beta}  - f_{k'}^{*}f_{k} \xi(\omega_{0},\omega_{k'},\omega_{k},\tau)\,\mathrm{e}^{-i(\omega_{k}-\omega_{k'})\tau}\,\varrho_{B}^{\beta}\mathrm{a}^{\dag}_{k}\mathrm{a}_{k'}\Big\}\nonumber\\
&+\mathrm{Tr}[\varrho_{S}(0)\sigma_{+}\sigma_{-}]\sum_{kk'}\Big\{f_{k}f^{*}_{k'}\,\eta(\omega_{0},\omega_{k},\tau)\eta^{*}(\omega_{0},\omega_{k'},\tau)\,\mathrm{a}^{\dag}_{k}\,\varrho_{B}^{\beta}\,\mathrm{a}_{k'}\nonumber\\
& -f^{*}_{k'}f_{k}\xi(\omega_{0},\omega_{k'},\omega_{k},\tau)\,\mathrm{e}^{-i\tau(\omega_{k}-\omega_{k'})}\,\mathrm{a}_{k'}\mathrm{a}^{\dag}_{k}\,\varrho_{B}^{\beta}  - f_{k'}f^{*}_{k} \xi^{*}(\omega_{0},\omega_{k'},\omega_{k},\tau)\,\mathrm{e}^{i(\omega_{k}-\omega_{k'})\tau}\,\varrho_{B}^{\beta}\mathrm{a}_{k}\mathrm{a}^{\dag}_{k'}\Big\}\Big)+O(\lambda^3), \label{rhob-2}
\end{align}
where 
\begin{align}
\eta(\omega_0,\omega_k,\tau) =& \int_{0}^{\tau}\mathrm{d}s\,\mathrm{e}^{i(\omega_0-\omega_k)s},
\label{eta-eq}
\\
\xi(\omega_0,\omega_{k'},\omega_{k},\tau) =& \int_{0}^{\tau}\mathrm{d}s_{1}\,\mathrm{e}^{i(\omega_0 -\omega_{k'})s_{1}}\eta^{*}(\omega_0,\omega_{k},s_{1}), \label{xi-eq}
\end{align}
$\overline{n}(\omega,\beta)$ shows the Planck distribution or the mean quanta number in a mode with frequency $\omega$ [Eq.~(\ref{def:n})], and $\varrho_{S}^{(0)}(\tau) = U_{S}(\tau)\varrho_{S}(0)U_{S}^{\dag}(\tau)$ is the unperturbed state of $S$, in which $U_{S}(\tau)=\mathrm{e}^{-i\tau H_{S}}$ (with $H_S=\omega_{0}\sigma_{z}/2$) is the free-system evolution.

In the continuum-$\omega$ limit, $\sum_{k}\to\int_{0}^{\infty}\mathrm{d}\omega$, we can find the dynamical equation of $\varrho_{S}^{(\lambda)}(\tau)$. We differentiate the continuum version of Eq.~(\ref{rhos-2}) in which we take $\tau\to\infty$ in the integrals of the RHS (long-time limit). In the long-time, weak-coupling limit we have $\tau\to\infty$ and $\lambda\to 0$ such that $\lambda^{2}\tau=\mathrm{const}$. This differentiation yields the Lindblad-type equation (\ref{lindblad-thermal-2}).

\subsection{Calculating thermodynamical properties}

Using the following notation for the states of the system and the bath:
\begin{align}
{\varrho}^{(\lambda)}_{S}(\tau) &= \varrho_{S}^{(0)}(\tau) + \lambda^2 \varrho^{(2)}_{S}(\tau) + O(\lambda^3),\\
{\varrho}^{(\lambda)}_{B}(\tau)&= \varrho_{B}^{\beta}+ \lambda \varrho^{(1)}_{B}(\tau) + \lambda^2 \varrho^{(2)}_{B}(\tau)+ O(\lambda^3), \label{rho-lambda-b}
\end{align}
the effective Hamiltonians of $S$ and $B$ can be computed up to $O(\lambda^3)$ as
\begin{align}
H_{S}^{(\mathrm{eff})}(\tau)&=H_{S}+\lambda\, \mathrm{Tr}_{B}\left[\varrho^{(1)}_{B}(\tau)\,H_{\mathrm{int}}^{(\lambda)}\right]
-\lambda \alpha_{S}\mathrm{Tr}\left[\varrho_{S}^{(0)}(\tau)\otimes \varrho^{(1)}_{B}(\tau)\,H_{\mathrm{int}}^{(\lambda)}\right], \label{h-eff-s}\\
H_{B}^{(\mathrm{eff})}(\tau)&= H_{B} + \mathrm{Tr}_{S}\left[\varrho^{(0)}_{S}(\tau)\,H_{\mathrm{int}}^{(\lambda)}\right]- \lambda \alpha_{B}\mathrm{Tr}\left[\varrho_{S}^{(0)}(\tau)\otimes \varrho^{(1)}_{B}(\tau)\,H_{\mathrm{int}}^{(\lambda)}\right].
\end{align}
We obtain
\begin{align*}
\mathrm{Tr}_{B}\left[\varrho^{(1)}_{B}(\tau)\,H_{\mathrm{int}}^{(\lambda)}\right]&= 2\lambda \sum_{k}|f_{k}|^{2} \left( i \varrho_{10}\, \sigma_{-}\int_{0}^{\tau}\mathrm{d}s\,\mathrm{e}^{i \omega_k \tau} \mathrm{e}^{i (\omega_0-\omega_k)s} + \mathrm{h.c.} \right),\\
\mathrm{Tr}_{S}\left[\varrho^{(0)}_{S}(\tau)\,H_{\mathrm{int}}^{(\lambda)}\right]& =2 \lambda \sum_{k} \left( f^{*}_k \varrho_{10} \mathrm{e}^{i \omega_0 \tau} \mathrm{a}_{k} + \mathrm{h.c.} \right) =: \lambda H_{B}^{(1)}(\tau),\\
\mathrm{Tr}\left[\varrho_{S}^{(0)}(\tau)\otimes \varrho^{(1)}_{B}(\tau)\,H_{\mathrm{int}}^{(\lambda)}\right] &= 8 \lambda |\varrho_{10}|^{2} \sum_{k}|f_{k}|^{2}\frac{1-\cos[(\omega_0-\omega_k)\tau]}{(\omega_0-\omega_k)} =: \lambda H_{B}^{(2)}(\tau),
\end{align*}
where ``$\mathrm{h.c.}$" denotes Hermitian conjugate. The energy of the bath then becomes
\begin{align}
\mathds{U}^{(\lambda)}_B(\tau)&= \mathrm{Tr}\left[\varrho^{(\lambda)}_{B}(\tau)\,H_{B}^{(\mathrm{eff})}(\tau)\right]\nonumber \\ 
&=\mathds{U}^{(0)}_B + \lambda^2 \left(-\alpha_{B} \mathrm{Tr}\left[\varrho_{B}^{\beta}\,H_{B}^{(2)}(\tau)\right]+ \mathrm{Tr}\left[\varrho_{B}^{(1)}(\tau)\,H_{B}^{(1)}(\tau)\right] + \mathrm{Tr}\left[\varrho_{B}^{(2)}(\tau)\,H_{B}\right] \right)+O(\lambda^3), 
\end{align}
which gives
\begin{align}
\mathrm{d}\mathds{U}^{(\lambda)}_{B}(\tau)=& \mathrm{d}\mathds{Q}_{B}(\tau)+ \mathrm{d}\mathds{W}_{B}(\tau) \nonumber \\
=& \lambda^2 \left( \mathrm{Tr}\left[\mathrm{d}\varrho_{B}^{(2)}(\tau)\,H_{B}\right] + \mathrm{Tr}\left[\mathrm{d}\varrho_{B}^{(1)}(\tau)\,H_{B}^{(1)}(\tau)\right]+ \mathrm{Tr}\left[\varrho_{B}^{(1)}(\tau)\,\mathrm{d}H_{B}^{(1)}(\tau)\right] -\alpha_{B} \mathrm{Tr}\left[\varrho_{B}^{\beta}\,\mathrm{d} H_{B}^{(2)}(\tau)\right]  \right)+O(\lambda^3).
\end{align}
After some straightforward algebra we can see that
\begin{align}
\mathrm{Tr}\left[\mathrm{d}\varrho_{B}^{(2)}(\tau)\,H_{B}\right]=& 8 \left[\big( \overline{n}(\omega_{k},\beta)+1\big)\varrho_{00} -\overline{n}(\omega_{k},\beta) \varrho_{11} \right] \sum_{k}|f_{k}|^{2} \frac{\omega_k}{(\omega_0-\omega_k)} \sin[(\omega_0-\omega_k)\tau]\,\mathrm{d}\tau, \label{drhohb}\\
\mathrm{Tr}\left[\varrho_{B}^{(1)}(\tau)\,\mathrm{d}H_{B}^{(1)}(\tau)\right]=& 8 |\varrho_{10}|^{2} \sum_{k}|f_{k}|^{2} \frac{\omega_0}{(\omega_0-\omega_k)} \sin[(\omega_0-\omega_k)\tau]\,\mathrm{d}\tau,\\
\mathrm{Tr}\left[\mathrm{d}\varrho_{B}^{(1)}(\tau)\,H_{B}^{(1)}(\tau)\right]=& -8 |\varrho_{10}|^{2} \sum_{k}|f_{k}|^{2} \frac{\omega_k}{(\omega_0-\omega_k)} \sin[(\omega_0-\omega_k)\tau]\,\mathrm{d}\tau,\\
\mathrm{Tr}\left[\varrho_{B}^{\beta}\,\mathrm{d} H_{B}^{(2)}(\tau)\right]=& 8 |\varrho_{10}|^{2} \sum_{k}|f_{k}|^{2} \sin[(\omega_0-\omega_k)\tau]\,\mathrm{d}\tau.
\end{align}
Hence 
\begin{equation}
\mathrm{d}\mathds{U}^{(\lambda)}_{B}(\tau)= 8\lambda^2 \sum_{k}|f_{k}|^{2} \sin[(\omega_0-\omega_k)\tau] \left[|\varrho_{10}|^{2} (1-\alpha_B) + \frac{\omega_k}{(\omega_0-\omega_k)}\left[\big( \overline{n}(\omega_{k},\beta)+1\big)\varrho_{00} -\overline{n}(\omega_{k},\beta) \varrho_{11} \right] \right]\,\mathrm{d}\tau +O(\lambda^{3}).
\label{ex:duB}
\end{equation}

For the entropy we have
\begin{equation}
\mathrm{d}\mathds{S}^{(\lambda)}_B(\tau)= - \mathrm{Tr}\left[\mathrm{d}\varrho_B^{(\lambda)}(\tau)\,\log\varrho_B^\beta\right] -\mathrm{Tr}\left[\mathrm{d}\varrho^{(\lambda)}_B(\tau)\,\left(\log\varrho^{(\lambda)}_B(\tau)\,-\,\log\varrho_B^\beta\right)\right],
\label{dsB}
\end{equation}
where the first term has already been computed as
\begin{align}
-\mathrm{Tr}\left[\mathrm{d}\varrho^{(\lambda)}_B(\tau)\,\log\varrho_B^\beta\right]=& \lambda^2 \beta\, \mathrm{Tr}\left[\mathrm{d}\varrho_{B}^{(2)}(\tau)\,H_{B}\right] + O(\lambda^3)\nonumber\\
\overset{\mathrm{(\ref{drhohb})}}{=}& 8\lambda^{2} \beta \Big[\big( \overline{n}(\omega_{k},\beta)+1\big)\varrho_{00} -\overline{n}(\omega_{k},\beta) \varrho_{11} \Big] \sum_{k}|f_{k}|^{2} \frac{\omega_k}{(\omega_0-\omega_k)} \sin[(\omega_0-\omega_k)\tau]\,\mathrm{d}\tau +O(\lambda^3).
\end{align}
In order to evaluate the second term of Eq.~(\ref{dsB}) we only need to take care of the contribution of order $\lambda$. We use the following integral form for the logarithm of an operator \cite{book:Hiai}:
\begin{equation}
\log A=\int_{0}^{\infty}\mathrm{d}x\,\left[ \frac{\openone}{1+x}-(x\openone+A)^{-1}\right],
\label{def:log}
\end{equation}
to obtain
\begin{align}
\log\varrho^{(\lambda)}_B(\tau)\,-\,\log\varrho_B^\beta=&\int_0^{\infty}\mathrm{d}x\,\left[\big(x\openone+ \varrho_{B}^{\beta}\big)^{-1}-\big(x\openone + \varrho^{(\lambda)}_B(\tau)\big)^{-1}\right]\nonumber\\
=& \lambda\,\int_0^{\infty}\mathrm{d}x\,\big(x\openone+\varrho_{B}^{\beta}\big)^{-1}\,\varrho^{(1)}_{B}(\tau)\, \big(x\openone+\varrho_{B}^{\beta}\big)^{-1}+\,O(\lambda^2),
\end{align}
where we have used the identity \cite{book:Hiai}
\begin{equation}
(A+B)^{-1} = A^{-1} - A^{-1}B A^{-1} + A^{-1}BA^{-1}B A^{-1} -O(B^3)
\end{equation}
to write
\begin{equation*}
\big(x\openone+\varrho_B^{(\lambda)}(\tau)\big)^{-1}=\big(x\openone+\varrho_{B}^{\beta}\big)^{-1}+\big(x\openone+\varrho_{B}^{\beta}\big)^{-1} \left(\varrho_B^\beta-\varrho^{(\lambda)}_B(\tau)\right)\big(x\openone+\varrho_{B}^{\beta}\big)^{-1} +O(\lambda^2)
\end{equation*}
and Eq.~(\ref{rho-lambda-b}).

To ease notation, we introduce $O_\tau=\mathrm{a}^\dag(h_\tau)-\mathrm{a}(h_\tau)$, with
\begin{equation*}
\mathrm{a}(h_\tau)= i \varrho_{10} \sum_{k}f^{*}_{k}\mathrm{e}^{i\omega_{k}\tau}\eta(\omega_{0},\omega_{k},\tau)\,\mathrm{a}_{k} ,
\end{equation*}
where we have followed the shorthand introduced in Eq.~(\ref{notation:af}) to define the vector $h_{\tau}=\{h_{k}(\tau)\}$, with $h_{k}(\tau) = -i\varrho^{*}_{10}f_{k}\mathrm{e}^{-i\omega_{k}\tau}\eta^{*}(\omega_{0},\omega_{k},\tau)$. Thus we can rewrite $\varrho^{(1)}_{B}(\tau)$ as
\begin{equation}
\varrho^{(1)}_{B}(\tau)= [O_\tau\,,\,\varrho_B^\beta],
\end{equation}
whence
\begin{equation}
-\mathrm{Tr}\left[\mathrm{d}\varrho^{(\lambda)}_B(\tau)\,\left(\log\varrho^{(\lambda)}_B(\tau)\,-\,\log\varrho_B^\beta\right)\right]=-\lambda^2\,\int_0^{\infty}\mathrm{d}x\,\mathrm{Tr}\left[[\mathrm{d}O_\tau\,,\,\varrho_B^\beta]\,\big(x\openone+\varrho_{B}^{\beta}\big)^{-1}\big[O_\tau\,,\,\varrho_B^\beta\big]\big(x\openone+\varrho_{B}^{\beta}\big)^{-1}\right]+O(\lambda^3).
\label{auxvNeb4}
\end{equation}
Considering the spectral decomposition $\varrho_B^\beta=\sum_{n}r_{n}|n\rangle\langle n|$, one can see
\begin{align}
\mathrm{Tr}\left[[\mathrm{d}O_\tau\,,\,\varrho_B^\beta]\,\big(x\openone+\varrho_{B}^{\beta}\big)^{-1}[O_\tau\,,\,\varrho_B^\beta]\big(x\openone+\varrho_{B}^{\beta}\big)^{-1}\right]=-\sum_{n,m}\langle n\vert\mathrm{d}O_\tau\vert m\rangle\langle m \vert O_\tau\vert n\rangle\,
\frac{(r_n-r_m)^2}{(x+r_n)(x+r_m)},
\end{align}
which yields
\begin{align}
\nonumber
\hskip-1cm
\int_0^{\infty}\mathrm{d}x\,\mathrm{Tr}\left[[\mathrm{d}O_\tau\,,\,\varrho_B^\beta]\,\big(x\openone+\varrho_{B}^{\beta}\big)^{-1}[O_\tau\,,\,\varrho_B^\beta]\big(x\openone+\varrho_{B}^{\beta}\big)^{-1}\right]=&\sum_{n,m}(r_m-r_n)\log\frac{r_n}{r_m}\,\langle n\vert\mathrm{d}O_\tau\vert m\rangle\langle m\vert O_\tau\vert n\rangle\\
\nonumber
=&\mathrm{Tr}\left[\varrho_B^\beta\left(\left[O_\tau\,,\,\log\varrho_B^\beta\right]\mathrm{d}O_\tau+\left[\mathrm{d}O_\tau\,,\,\log\varrho_B^\beta\right]O_\tau\right)\right]\\
\nonumber
=&2\beta\sum_k\omega_k\,\mathrm{Re}\left[h_k(\tau)\,\mathrm{d}h^*_k(\tau)\right]\\
\nonumber
=&8\beta|\varrho_{10}|^{2} \sum_{k}|f_{k}|^{2} \frac{\omega_k}{(\omega_0-\omega_k)} \sin[(\omega_0-\omega_k)\tau]\,\mathrm{d}\tau.
\end{align}
Thus, noting Eq.~(\ref{dsB}), we obtain
\begin{equation}
\mathrm{d}\mathds{S}^{(\lambda)}_B(\tau)= 8\lambda^2 \beta \sum_{k}|f_{k}|^{2}\frac{\omega_k \sin[(\omega_0-\omega_k)\tau]}{(\omega_0-\omega_k)}  \Big[  \big( \overline{n}(\omega_{k},\beta)+1\big)\varrho_{00} -\overline{n}(\omega_{k},\beta) \varrho_{11}  - |\varrho_{10}|^{2} \Big]\mathrm{d}\tau +O(\lambda^{3}).
\label{ex:dsB}
\end{equation}
Now combining Eqs.~(\ref{ex:duB}) and (\ref{ex:dsB}), the pseudo-temperature $T_{B}^{(\lambda)}(\tau)$ reads as
\begin{align}
T_{B}^{(\lambda)}(\tau) &=\frac{\mathrm{d}\mathds{U}^{(\lambda)}_{B}(\tau)}{\mathrm{d}\mathds{S}^{(\lambda)}_{B}(\tau)} \nonumber\\
&= \frac{1}{\beta} \frac{ \sum_{k}|f_{k}|^{2} \frac{\omega_{k}\sin[(\omega_0-\omega_k)\tau]}{(\omega_0-\omega_k)}  \Big[  \big( \overline{n}(\omega_{k},\beta)+1\big)\varrho_{00} -\overline{n}(\omega_{k},\beta) \varrho_{11} - |\varrho_{10}|^{2} + |\varrho_{10}|^{2}[\alpha_{B}(\omega_{k}-\omega_{0}) +\omega_{0}]/\omega_{k}\Big]}{ \sum_{k}|f_{k}|^{2}\frac{ \omega_{k}\sin[(\omega_0-\omega_k)\tau]}{(\omega_0-\omega_k)}  \Big[  \big( \overline{n}(\omega_{k},\beta)+1\big)\varrho_{00} -\overline{n}(\omega_{k},\beta) \varrho_{11} - |\varrho_{10}|^{2} \Big] }.
\end{align}

If we go to the continuum-$\omega$ limit, take the $\tau\to\infty$ limit, and use the identity
\begin{equation}
\lim_{\tau\to\infty}\frac{\sin(x\tau)}{\pi x}=\delta(x),
\end{equation}
we obtain
\begin{align}
\lim_{\tau\to\infty}T_{B}^{(\lambda)}(\tau) &= \frac{1}{\beta} \frac{ \Big[  \big( \overline{n}(\omega_{0},\beta)+1\big)\varrho_{00} -\overline{n}(\omega_{0},\beta) \varrho_{11} \Big]}{ \Big[  \big( \overline{n}(\omega_{0},\beta)+1\big)\varrho_{00} -\overline{n}(\omega_{0},\beta) \varrho_{11} - |\varrho_{10}|^{2} \Big]}\nonumber\\
&= \frac{1}{\beta} \Big[ 1+\frac{ |\varrho_{10}|^{2} }{ \overline{n}(\omega_{0},\beta) (\varrho_{00} - \varrho_{11})+\varrho_{00} - |\varrho_{10}|^{2} }\Big].
\label{tb-limit-app}
\end{align}

Let us now study system $S$. Since we are interested in thermalization we consider the solution to the Lindblad equation \eqref{lindblad-thermal-2}, which is given by 
\begin{equation}
\varrho^{(\lambda)}_{S}(\tau)= \frac{1}{2} 
    \begin{bmatrix}
    &1+ z(0)\mathrm{e}^{- \widetilde{\gamma}\tau}+ \tanh(\beta \omega_0 /2)\left( \mathrm{e}^{- \widetilde{\gamma}\tau} -1 \right)  & \big(x(0)- i y(0)\big)\, \mathrm{e}^{- \widetilde{\gamma}\tau/2-i\omega_0 \tau}  \\
    &\big(x(0)+ i y(0)\big)\, \mathrm{e}^{- \widetilde{\gamma}\tau/2+i\omega_0 \tau}   & 1- z(0)\mathrm{e}^{- \widetilde{\gamma}\tau}- \tanh(\beta \omega_0 /2)\left( \mathrm{e}^{- \widetilde{\gamma}\tau} -1 \right)
    \end{bmatrix},
\label{solution-app}
    \end{equation}
where $\widetilde{\gamma}= \gamma \coth(\beta \omega_0 /2) $ and $\big(x(0),y(0),z(0)\big)$ are the initial components of the Bloch vector.  We can explicitly compute $\mathrm{d}\mathds{S}^{(\lambda)}_S(\tau)$ using the eigenvalues of $\varrho^{(\lambda)}_{S}(\tau)$, $(1/2)\big(1\pm \sqrt{x^2(\tau)+y^2(\tau)+z^2(\tau)}\big)$, as
\begin{align}
\mathrm{d}\mathds{S}^{(\lambda)}_S(\tau)&=-\frac{1}{2}\log\left( \frac{1+\sqrt{x^2(\tau)+y^2(\tau)+z^2(\tau)}}{1-\sqrt{x^2(\tau)+y^2(\tau)+z^2(\tau)}}\right)\mathrm{d}\left( \sqrt{x^2(\tau)+y^2(\tau)+z^2(\tau)} \right)\nonumber\\
&=-\frac{1}{2}\log\left( \frac{1+\sqrt{x^2(\tau)+y^2(\tau)+z^2(\tau)}}{1-\sqrt{x^2(\tau)+y^2(\tau)+z^2(\tau)}}\right)\frac{\frac{\widetilde{\gamma}}{2}\left(x^2(\tau)+y^2(\tau)\right)-\gamma z(\tau)-\widetilde{\gamma}z^2(\tau)}{\sqrt{x^2(\tau)+y^2(\tau)+z^2(\tau)}}\mathrm{d}\tau .
\end{align}
The energy of this system is 
\begin{align}
\mathds{U}^{(\lambda)}_S(\tau)&=\mathrm{Tr}\left[ \varrho^{(\lambda)}_{S}(\tau) H_{S}^{(\mathrm{eff})}(\tau)\right]\nonumber\\
&= \frac{\omega_0}{2} \mathrm{Tr}\left[ \varrho^{(\lambda)}_{S}(\tau) \sigma_z \right] + \lambda \left( 1-\alpha_S \right)\left(\mathrm{Tr}\left[  \varrho^{(\lambda)}_{S}(\tau) \sigma_+ \right]\mathrm{Tr}\left[  \varrho^{(\lambda)}_{B}(\tau) \mathrm{a}(f) \right]+\mathrm{Tr}\left[  \varrho^{(\lambda)}_{S}(\tau) \sigma_- \right]\mathrm{Tr}\left[  \varrho^{(\lambda)}_{B}(\tau) \mathrm{a}^{\dag}(f) \right] \right) \nonumber \\
&= \frac{\omega_0}{2}z(\tau)+ 2\lambda^2 \left( 1-\alpha_S \right)(x^2(0)+y^2(0))\mathrm{e}^{- \widetilde{\gamma}\tau}\sum_{k}|f_{k}|^{2}\frac{1-\cos[(\omega_0-\omega_k)\tau]}{(\omega_0-\omega_k)} + O(\lambda^3),
\end{align}
where we used Eq. \eqref{solution} for $\varrho^{(\lambda)}_{S}(\tau)$ and $\varrho^{(\lambda)}_{B}(\tau)=\varrho_{B}^{\beta}+\lambda\varrho^{(1)}_{B}(\tau) + O(\lambda^2)$.
Recalling Eq.~(\ref{gamma-def}), the expression above can be differentiated as follows:
\begin{align}
\mathrm{d}\mathds{U}^{(\lambda)}_S(\tau)=&  -\frac{\omega_0}{2}\gamma \mathrm{e}^{- \widetilde{\gamma}\tau}\Big( \coth(\beta \omega_0 /2)z(0) + 1\Big) \mathrm{d}\tau + \frac{\gamma \left( 1-\alpha_S \right)}{\pi |f(\omega_{0})|^{2}} (x^2(0)+y^2(0))\mathrm{e}^{- \widetilde{\gamma}\tau}\sum_{k}|f_{k}|^{2}\sin[(\omega_0-\omega_k)\tau]\,\mathrm{d}\tau\nonumber\\
\overset{\omega-\mathrm{continuum}}{=}& -\frac{\omega_0}{2}\gamma \mathrm{e}^{- \widetilde{\gamma}\tau}\Big( \coth(\beta \omega_0 /2)z(0) + 1\Big) \mathrm{d}\tau.
\end{align}
As a result, the inverse pseudo-temperature becomes
\begin{align}
\frac{1}{T^{(\lambda)}_{S}(\tau)} =& -\frac{1}{2}\log\left( \frac{1+\sqrt{x^2(\tau)+y^2(\tau)+z^2(\tau)}}{1-\sqrt{x^2(\tau)+y^2(\tau)+z^2(\tau)}}\right)\frac{\frac{1}{2}\coth(\beta \omega_0 /2)\big(x^2(0)+y^2(0)\big)+ z(\tau)\Big( \coth(\beta \omega_0 /2)z(0) + 1\Big)}{\sqrt{x^2(\tau)+y^2(\tau)+z^2(\tau)}(\omega_0/2)\Big( \coth(\beta \omega_0 /2)z(0) + 1\Big)},
\end{align}
which yields
\begin{equation}
\lim_{\tau\to\infty}\frac{1}{T^{(\lambda)}_{S}(\tau)} =\beta  \left[ 1 - \frac{\big(x^2(0)+y^2(0)\big)\coth(\beta \omega_0 /2)}{2\big(z(0)+ \tanh(\beta \omega_0 /2)\big)} \right].
\end{equation}
Thus, similarly to the case of $\lim_{\tau\to 0}T^{(\lambda)}_{B}(\tau)$, in this case too the pseudo-temperature $T^{(\lambda)}_{S}(\tau)$ behaves as expected if there is no initial coherence ($\varrho_{10}=0$, or equivalently, $x(0)=y(0)=0$). 

\section{Details of example II}
\label{app:2}

If we expand $\varrho_{B}^{(\lambda)}(\tau) =: \varrho_{B}^{\beta} + \lambda \varrho^{(1)}_{B}(\tau) + \lambda^2 \varrho^{(2)}_{B}(\tau) +O(\lambda^3)$, we obtain
\begin{align}
\varrho^{(1)}_{B}(\tau)= & \langle \sigma_z\rangle_{S}\Big[ \sum_{k}\big( g_k(\tau) \mathrm{a}^{\dag}_{k} - g^{*}_k(\tau) \mathrm{a}_{k} \big),\varrho_{B}^{\beta}\Big], \label{R1}\\
\varrho^{(2)}_{B}(\tau) =& (1/2)\Big\{ \sum_{kk'}\big( g_k(\tau) \mathrm{a}^{\dag}_{k} - g^{*}_k(\tau) \mathrm{a}_{k} \big)\big( g_{k'}(\tau) \mathrm{a}^{\dag}_{k'} - g^{*}_{k'}(\tau) \mathrm{a}_{k'} \big),\varrho_{B}^{\beta}\Big\} \nonumber\\
&-\sum_{k}\big( g_k(\tau) \mathrm{a}^{\dag}_{k} - g^{*}_k(\tau) \mathrm{a}_{k} \big) \varrho_{B}^{\beta} \sum_{k'}\big( g_{k'}(\tau) \mathrm{a}^{\dag}_{k'} - g^{*}_{k'}(\tau) \mathrm{a}_{k'} \big).
\label{R2}
\end{align}
Since we need to compute the entropy $\mathds{S}_{B}^{(\lambda)}(\tau)=-\mathrm{Tr}[\varrho^{(\lambda)}_{B}(\tau)\log\varrho^{(\lambda)}_{B}(\tau)]$, we shall need to calculate $\log\varrho_{B}^{(\lambda)}(\tau)$ up to $O(\lambda^{3})$. In order to do so, we use the following identity \cite{book:Hiai}:
\begin{align}
\log(A_0 + \lambda A_1 + \lambda^2 A_2) =& \log A_0 + \lambda \int_{0}^{\infty}\mathrm{d}x\,(A_0 +x\openone)^{-1} A_{1} (A_0 + x\openone)^{-1} \nonumber\\
&-\lambda^{2} \int_{0}^{\infty} \mathrm{d}x\,\Big[ (A_0 +x\openone)^{-1} A_{1} (A_0 +x\openone)^{-1} A_{1} (A_0 +x\openone)^{-1} - (A_0 +x\openone)^{-1} A_{2} (A_0 +x\openone)^{-1}\Big]\nonumber\\
&+ O(\lambda^3) =: L_0 + \lambda L_1 + \lambda^2 L_2 + O(\lambda^3). 
\label{log-expansion}
\end{align}
Replacing the terms of $\varrho_{B}^{(\lambda)}(\tau)$ in Eq.~(\ref{log-expansion}) yields
\begin{align}
L_0 & = \log \varrho_{B}^{\beta},\label{L0}\\
L_1(\tau) &= \beta \langle \sigma_z\rangle_0 \sum_{k}\omega_{k} \big( g_k(\tau) \mathrm{a}^{\dag}_{k} + g^{*}_k(\tau) \mathrm{a}_{k} \big).
\end{align}
Hence
\begin{align}
\mathds{S}_{B}^{(\lambda)}(\tau) =& - \mathrm{Tr}\Big[\Big( \varrho_{B}^{\beta} + \lambda \varrho^{(1)}_{B}(\tau)+ \lambda^2 \varrho^{(2)}_{B}(\tau)\Big) \Big(L_0 + \lambda L_{1}(\tau)+ \lambda^2 L_{2}(\tau) \Big)\Big]+O(\lambda^3)\nonumber\\
=& -\mathrm{Tr}[\varrho_{B}^{\beta} L_0] - \lambda\Big(\mathrm{Tr}[\varrho_{B}^{\beta} L_1(\tau)] + \mathrm{Tr}[\varrho^{(1)}_{B}(\tau) L_0] \Big) -\lambda^2 \Big( \mathrm{Tr}[\varrho_{B}^{\beta} L_2(\tau)]+ \mathrm{Tr}[\varrho^{(1)}_{B}(\tau) L_1(\tau)] + \mathrm{Tr}[\varrho^{(2)}_{B}(\tau) L_0] \Big)\nonumber\\
& + O(\lambda^3).
\end{align}
From this relation we obtain
\begin{align}
\mathrm{d}\mathds{S}_{B}^{(\lambda)}(\tau) =& - \lambda\Big(\mathrm{Tr}[\varrho_{B}^{\beta}\, \mathrm{d}L_1(\tau)] + \mathrm{Tr}[\mathrm{d}\varrho^{(1)}_{B}(\tau)\, L_0] \Big) -\lambda^2 \Big( \mathrm{Tr}[\varrho_{B}^{\beta} \,\mathrm{d}L_2(\tau)]+ \mathrm{Tr}[\mathrm{d}\varrho^{(1)}_{B}(\tau) \,L_1(\tau)] + \mathrm{Tr}[\varrho^{(1)}_{B}(\tau)\, \mathrm{d}L_1(\tau)]\nonumber\\&+ \mathrm{Tr}[\mathrm{d}\varrho^{(2)}_{B}(\tau)\, L_0] \Big) + O(\lambda^3).
\end{align}
This expression has some irrelevant (i.e., vanishing) terms. This can be seen through the identity $\mathrm{d}\mathds{S}(\tau)=-\mathrm{Tr}[\mathrm{d}\varrho\log\varrho]$, from whence
\begin{align}
\mathrm{d}\mathds{S}_{B}^{(\lambda)}(\tau) =& - \lambda\mathrm{Tr}[\mathrm{d}\varrho^{(1)}_{B}(\tau)\, L_0] -\lambda^2 \Big( \mathrm{Tr}[\mathrm{d}\varrho^{(1)}_{B}(\tau) \,L_1(\tau)] + \mathrm{Tr}[\mathrm{d}\varrho^{(2)}_{B}(\tau)\, L_0] \Big)+ O(\lambda^3). \label{ds-1}
\end{align}
One can see from the identity $\mathrm{Tr}\big[[A,B]f(B) \big]=0$ (for any $A$, $B$, and function $f$) that here
\begin{equation}
\mathrm{Tr}[\mathrm{d}\varrho^{(1)}_{B}(\tau)\, L_0] \overset{\mathrm{(\ref{R1}),~(\ref{L0})}}{=}0.
\end{equation}
 Thus Eq. (\ref{ds-1}) reduces to
\begin{align}
\mathrm{d}\mathds{S}_{B}^{(\lambda)}(\tau) &= -\lambda^2 \Big( \mathrm{Tr}[\mathrm{d}\varrho^{(1)}_{B}(\tau) \,L_1(\tau)] + \mathrm{Tr}[\mathrm{d}\varrho^{(2)}_{B}(\tau)\, L_0] \Big)+ O(\lambda^3),\nonumber\\
&= 4\beta\lambda^2(1-\langle\sigma_z\rangle_S^2)\,\mathrm{d}\Delta(\tau).
\end{align}

\twocolumngrid
\end{widetext}


\end{document}